\documentclass[a4paper,11pt]{article}
\usepackage{jcappub} 
\newcommand{\afterETSpace}{}
\usepackage{lineno}

\usepackage{letltxmacro}

\LetLtxMacro{\oldautoref}{\autoref}
\renewcommand{\autoref}[1]{\textsc{\oldautoref{#1}}}

\usepackage{xcolor}
\usepackage{float}
\usepackage{hyperref}
\definecolor{sapred}{rgb}{0.5198039,0.1411765,0.2}
\hypersetup{
colorlinks=true,
linkcolor=sapred,
citecolor=sapred,
urlcolor=sapred
}
\usepackage{multirow}
\usepackage{xfrac,bigints}
\usepackage{booktabs}
\usepackage{tabularx}
\usepackage{ragged2e}

\def\cdlt#1{\textcolor{black}{#1}} 

\ETnumber{\, ET-0612A-25}

\title{Illuminating the Dark Sector: Understanding Modified Gravity Signatures with Cross-Correlations of Gravitational Waves and Large-Scale Structure}

\author[a,b,c,d]{C. De Leo,}
\author[e,f,g]{G. Cañas-Herrera,}
\author[g,h]{A. Balaudo,}  
\author[b,c]{M. Martinelli,}
\author[i]{A. Silvestri,}
\author[h]{T. Baker}

\affiliation[a]{Sapienza University, Piazzale Aldo Moro, 2 - c/o Dipartimento di Fisica, Edificio E. Fermi, I-00185 Rome, Italy}
\affiliation[b]{INAF - Osservatorio Astronomico di Roma, via Frascati 33, 00040 Monteporzio Catone, Italy}
\affiliation[c]{INFN - Sezione di Roma, Piazzale Aldo Moro, 2 - c/o Dipartimento di Fisica, Edificio G. Marconi, I-00185 Rome, Italy}
\affiliation[d]{Tor Vergata University, Via della Ricerca Scientifica, 1 - c/o Dipartimento di Fisica, I-00133 Rome, Italy}
\affiliation[e]{Institute for Astronomy, School of Physics and Astronomy, University of Edinburgh, Royal Observatory, Blackford Hill, Edinburgh EH9 3HJ, United Kingdom}
\affiliation[f]{Leiden Observatory, Leiden University, Niels Bohrweg 2, 2333 CA, Leiden, The Netherlands}
\affiliation[g]{ESTEC - European Space Agency, Keplerlaan 1, 2201 AZ Noordwijk, The Netherlands}
\affiliation[h]{Institute of Cosmology and Gravitation, University of Portsmouth, Burnaby Road, Portsmouth PO1 3FX, United Kingdom}
\affiliation[i]{Institute Lorentz, Leiden University, PO Box 9506, Leiden 2300 RA, The Netherlands}

\emailAdd{chiara.deleo@uniroma1.it}

\abstract{We investigate the synergy between large-scale structure (LSS) observations and gravitational wave (GW) events for testing modified gravity. In particular, we forecast the LSS $\times$ GW cross-correlation signal using Stage-IV LSS surveys, such as Euclid, in combination with future detections from the Einstein Telescope. This cross-correlation provides a novel probe of fundamental physics, potentially revealing deviations from the $\Lambda$CDM paradigm that may not be accessible through electromagnetic observations alone. We describe the considered modified gravity scenarios, the relevant LSS and GW observables, and the synthetic forecast methodology. Our results demonstrate that combining LSS and GWs can significantly enhance constraints on departures from General Relativity, opening a new window for multi-messenger cosmology. We further assess the observational requirements GW experiments must meet to improve upon constraints obtainable from LSS alone.}

\begin{document}
\maketitle
\flushbottom

\section{Introduction}
\label{sec:intro}

The advances of  observational cosmology over the past decades have established $\Lambda$CDM as the standard cosmological model with cold dark matter (CDM) and  dark energy (DE), in the form of the cosmological constant $\Lambda$, as the main components. With only a handful of free parameters, the model has been very successful in matching different observations, from  the detailed observations of the Cosmic Microwave Background~\citep{WMAP,Planck18}, to large-scale structure (LSS) data by both photometric \citep{DESY3, KiDS2021_modelling}, and spectroscopic galaxy surveys \citep{eBOSS}.
With the advent of precision cosmology, some cracks have started to emerge, revealing tensions \citep{CosmoVerseNetwork:2025alb, Di_Valentino_2021, Verde_2019, Abdalla_review} and potential deviations \citep{DESIDR1, DESIDR2} that might hint at the need to extend beyond the standard model of cosmology with, for instance, dynamical DE or modifications of gravity (MG). 

Until now, the Universe has primarily been observed through the electromagnetic (EM) spectrum. However, the direct detection of gravitational waves (GW) by the 
LIGO-Virgo-KAGRA (LVK) collaboration has
opened a new observational window onto the cosmos. Additionally, the first detection of a GW with an electromagnetic counterpart \citep{2017ApJ...848L..13A}, marked the beginning of the era of multi-messenger analysis, allowing for an independent measurement of the Hubble constant \citep{Abbott:2017xzu} and a stringent, low-redshift bound on the speed of gravitational waves \citep{2017ApJ...848L..13A} which implied powerful constraints on the DE theoretical landscape \citep{Bettoni2017,Creminelli:2017sry,PhysRevLett.119.251304,PhysRevLett.119.251301}. Looking ahead, the next generation of ground-based GW detectors, including the Einstein Telescope (ET) \citep{Abac:2025saz} and Cosmic Explorer (CE) \citep{CE}, promises to significantly increase the sensitivity of these GW signals. These detectors have the capacity to measure hundreds of thousands of events over a decade, making statistical analyses for multi-messenger astrophysics a reality. This will eventually allow us to probe a wide range of cosmic phenomena, from binary black hole systems to massive black hole binaries and even a potential stochastic gravitational wave background \citep{Ca_as_Herrera_2020, LIGO_SGWB-2019}.

While next-generation LSS surveys, in combination with CMB data, will place very competitive constraints on cosmological models, the inclusion of GWs events, through cross-correlations with LSS, is promising to further increase the statistical constraining power and provide complementary information that can mitigate systematic effects present in individual datasets. By combining different probes, degeneracies between cosmological parameters can be broken and uncertainties reduced, allowing for a better understanding of possible tensions. For example, GW standard sirens can provide a direct measurement of the luminosity distance that is independent of the cosmic distance ladder, potentially helping to shed light on the Hubble tension. 
In general, this approach has already been explored in \citep{sala2025inferringcosmologicalparametersgalaxy, pedrotti2025cosmologyangularcrosscorrelationgravitationalwave, zazzera2025exploringfuturesynergieslargescale, Bosi_2023}.

In this work, we present a detailed forecast of the synergy between LSS and GW surveys, with focus on cross-correlations in the broad context of modified gravity. We adopt specifications from Stage-IV LSS surveys, such as the current ESA Euclid mission \citep{Laureijs11, EuclidSkyOverview}, and future data from the Einstein Telescope \citep{Abac:2025saz}. The LSS$\times$GW cross-correlation offers an unprecedented opportunity to test fundamental physics and expand our cosmological horizons beyond what is possible with EM observations alone, potentially providing a distinctive signature of extensions to $\Lambda$CDM.

The paper is organized as follows. We start describing the framework adopted to explore modified gravity in \autoref{sec:MG-DE}. In \autoref{sec:observables}, we present the different cosmological probes (\autoref{subsec:photo_observables}), GW observables (\autoref{subsec:gw_observables}) and the cross-correlations between the two (\autoref{subsec:cross_observables}). In \autoref{sec:methodology}, we introduce our methodology. 
Finally, we present and discuss the results in \autoref{sec:results} and draw main conclusions in \autoref{sec:conclusions}.                         

\section{Beyond-\texorpdfstring{$\Lambda$}{Lambda}CDM: phenomenological framework} 
\label{sec:MG-DE}

We adopt a phenomenological parametrization of modified gravity 
to explore potential deviations from General Relativity (GR) without restricting to specific models. In general, a MG theory will modify the dynamics of the background expansion and  LSS as well as  the propagation of GWs. 

Modifications to the background expansion history can be captured via the equation of state of dark energy, $w_{\rm DE}(z)$. In this work we focus on modifications to the evolution of perturbations, therefore we will keep the background fixed to that of $\Lambda$CDM, i.e. setting $w_{\rm DE}=-1$. Even when the background is fixed to $\Lambda$CDM, the luminosity distance associated to a GW event can generally be changed from its $\Lambda$CDM value due to two effects: i) the running of the Planck mass with redshift, and ii) the speed of propagation of GWs can differ from that of light. However, the tight constraints from GW170817 and its electromagnetic counterpart \citep{2017PhRvL.119p1101A} imply that the GW propagation speed is equal to the speed of light to very high precision, and we therefore set $c_T = c$ throughout this work.
Finally, the relativistic corrections to the luminosity distance and the number count of GWs can differ from that in $\Lambda$CDM \citep{Balaudo_2024}. We review all this in \autoref{app:GW_sources}. 
\newline
In screened scalar-tensor theories, local screening ensures that GW emission is governed by GR; accordingly, we neglect local modifications at emission and assume that deviations from the standard behaviour enter only through propagation effects encoded in the GW luminosity distance, as discussed in \autoref{app:GW_sources}.
Additionally, in the redshift range of interest for this work, the dominant relativistic correction will be the lensing one, which does not contain any explicit modifications. Still,  in  \autoref{app:GW_sources} we report all relativistic corrections, including those that contain explicit MG terms. The latter will be parametrized to broadly represent what is expected in the context of scalar-tensor theories of gravity. 

We are then left with modifications to the scalar sector only. 
We consider scalar perturbations around the Friedmann-Lemaitre-Robertson-Walker background metric in Newtonian gauge, with the perturbed line element given by \citep{Amendola_MGreview, GR_review}:
\begin{equation}
    ds^2=-(1+2\Psi)dt^2+a(t)^2(1-2\Phi)dx_idx^i,
    \label{eq:ds_MG}
\end{equation}
where $a(t)$ is the scale factor and $\Phi$ and $\Psi$ are the gravitational potentials \citep{Ma_1995}. Without loss of generality, it is possible to capture the effects of MG on LSS via  two phenomenological functions of time and scale introduced via the  Poisson equations for clustering and lensing
\citep{GR_review, Zhao_2009}:
\begin{equation}
\begin{aligned}
    -k^2 \Phi(a,k) &= 4\pi G a^2 \mu(a,k) \rho(a) \Delta(a,k), \\
    -k^2 \Psi_W(a,k) &= 4\pi G a^2 \Sigma(a,k) \rho(a) \Delta(a,k), \\
\end{aligned}
\label{eq:phi_and_psi}
\end{equation}
where $\Delta(a,k)\,=\,\delta(a,k)+3aH(a)\theta$ is the comoving density contrast of matter,  $\theta$ the peculiar velocity, $\Psi_W\equiv\frac{\Psi+\Phi}{2}$ the Weyl potential and we neglect anisotropic stress from neutrinos.
$\Lambda$CDM  is recovered for $\mu=\Sigma=1$\citep{GR_review}. 

For the purpose of forecasting the constraining power of the LSS$\times$GW, we will adopt the following common  parametrization for $\mu,\Sigma$, first introduced in \citep{Simpson_2012, Lee_2021}:
\begin{equation}
    \mu(z)\,=\,1\,+\,\mu_0\,\frac{\Omega_{\rm DE}(z)}{\Omega^0_{\rm DE}},
    \label{eq:mu}
\end{equation}
\begin{equation}
    \Sigma(z)\,=\,1\,+\,\Sigma_0\,\frac{\Omega_{\rm DE}(z)}{\Omega^0_{\rm DE}}.
    \label{eq:Sigma}
\end{equation}
In our case, $\Omega_{\rm DE}(z)=\Omega_\Lambda(z)=\rho_\Lambda/(3M_P^2H(z)^2)$ where $\rho_\Lambda$ is constant.

\section{Description of the cosmological observables}\label{sec:observables}
In this work we will consider photometric galaxy clustering (GC) and weak lensing (WL) as well as GW weak lensing (GW-WL) and number count (GW-NC). For all these observables, and the different cross-correlations, we will focus on their 2-point summary statistics in harmonic space, i.e. the angular power spectrum, after slicing the redshift interval in tomographic bins based on the photometric redshift for both galaxies and GWs. For these latter the natural observable is the luminosity distance, so we move from the luminosity distance D-space to redshift-space before defining the tomographic bins.

The angular power spectrum of any two observables A and B, correspondingly in the tomographic bin $i$ and $j$, has the general expression:
\begin{equation}
C^{\rm{A \times B}}_{ij} (\ell)= 4\pi \int d\log k \int_0^{\eta_0} d\eta W^A_i(\eta)j_\ell(k,d_C)\int_0^{\cdlt{\eta_0}} d\eta' W^B_j(\eta')j_\ell(k,d_C')P_{mm}(k,\eta,\eta'),
    \label{eq:C_ls_general}
\end{equation} 
where we have adopted the notation of \citep{Tessore2025}. Here $\eta$ is the conformal time, $j_\ell(k,d_C)$ are the Bessel functions and $d_C$ is the comoving distance;  $P_{mm}(k,\eta,\eta')$ is the matter power spectrum , which can be related to the primordial comoving curvature spectrum $P_\mathcal{R}(k)$ through the matter transfer function $T^2_{mm}(k,z)$: $P_{mm}(k,z) = T^2_{mm}(k,z)P_\mathcal{R}(k)$, where $P_\mathcal{R}(k) \propto A_s(k/k^*) ^{n_s - 1}$, with $A_s$ and $n_s$ the amplitude and spectral index, respectively. The window functions $W_i(z)$ are the corresponding probe kernels, and carry information about the experimental specifications and the underlying assumed cosmological model. We will present the details of these functions in the next subsections.

Eqs.~\eqref{eq:C_ls_general} can be simplified assuming both the Limber and the flat-sky approximations \citep{LoVerde_2008} that are valid for high multipoles $\ell$, typically $\ell>100$, where the spherical Bessel functions oscillate rapidly and can be averaged out, and the patches of the sky can be considered small so that the curvature is negligible. In this approximation the definition of the coefficients of angular power spectra become:
\begin{equation}
    C_{ij}^{\rm{A\times B}} (\ell) = \int dz \frac{c}{H(z)d_C^2(z)} W^A_i(z)W^B_j(z)P_{mm}\left(k(\ell,z),z\right),
    \label{eq:C_ls}
\end{equation}
where $k(\ell,z)=(\ell-1/2)/d_C(z)$. \cdlt{In this work, we adopt the Limber approximation in all computations of the angular power spectra.}

\subsection{Photometric Large-Scale Structure observables}\label{subsec:photo_observables}
For what concerns the LSS part, we consider weak gravitational lensing (WL), photometric angular galaxy clustering (GCph), and their cross-correlation. When combined, these observables constitute the photometric 3 $\times$ 2 pt analysis.

Gravitational lensing refers to the deflection of light by the intervening matter distribution. In the weak lensing regime, where the induced distortions are small, this effect can only be measured statistically through the coherent shape distortions of background galaxies, known as cosmic shear. WL is sensitive to the total matter content of the Universe, both baryonic and dark, and to the amplitude of matter-density fluctuations on large scales. On the other hand we know that the matter distribution in our Universe can be traced using galaxies. Through photometric galaxy clustering  we analyze how closely the large-scale distribution of galaxies can actually resemble the underlying matter distribution. In particular we know that these two are not identical, rather, they are biased due to the fact that galaxies do not perfectly trace the distribution of all matter, especially dark matter, but instead highlight denser regions where galaxy formation is more likely. Consequently, GCph can be used to get information about the matter distribution, discerning between matter components, the expansion rate of the Universe and also the growth rate of the structures. 

Keeping Eq.~\eqref{eq:C_ls} as a reference, we describe below how to define the quantities for each LSS observable given the corresponding window function $W(z)$. For the GCph the window function can be written as  \citep{Blanchard-EP7, CLOE1}:
\begin{equation}
    W^{\rm{GCph}}_i(z)=b^{\rm gal}_i n_i^{\rm gal}(z)\frac{H(z)}{c}\,,
    \label{window_gal}
\end{equation}
where:
\begin{itemize}
    \item $b^{\rm gal}_i$ is the bias parameter that quantifies how well galaxies trace the underlying matter distribution. In this work, we assume a polynomial bias \citep{EuclidSkyOverview}:
    \begin{equation}
    b^{\rm gal}(z)=b^{\rm gal}_0 +b^{\rm gal}_1z+b^{\rm gal}_2z^2 +b^{\rm gal}_3z^3,
    \label{bias}
    \end{equation}
    \item $n_i^{\rm gal}(z)$ is the galaxy redshift distribution that depends on the specifics of the survey, such as the geometry of the survey area, the redshift range covered, and the selection function that determines which galaxies are included in the sample and which are not.
\end{itemize}

For cosmic Weak Lensing, we need to take into account two contributions: shear signal $(\gamma)$ and the Intrinsic Alignment (IA). The latter summarizes the  contribution to the alignments of galaxies from local tidal forces, while $\gamma$ encodes the lensing contribution from structure along the line of sight. The angular power spectrum for WL can then be written as the sum of three contributions:
\begin{equation}
    C^{\rm{WL}}_{ij}(\ell)=C^{\gamma\gamma}_{ij}(\ell) + C^{\gamma\rm{IA}}_{ij}(\ell) + C^{\rm{IA}\rm{IA}}_{ij}(\ell).
    \label{Cls_WL}
\end{equation}
The window function for the shear is given by \citep{Blanchard-EP7}:
\begin{equation}
    W^{\gamma}(z)=\frac{3}{2} \Omega_m \frac{H_0^2}{c^2} d_C(z)(1+z)\int^\infty _z dx \; n_i^{\rm gal} \frac{d_C(x)-d_C(z)}{d_C(x)},
    \label{window_shear}
\end{equation}
where $\Omega_m$ is the matter density parameter. For IA we have \citep{Blanchard-EP7}:
\begin{equation}
    W^{\rm IA}(z)=f_{\rm IA}(z)\frac{H(z)n_{i}^{\rm gal}}{c},
    \label{window_IA}
\end{equation}
where we assumed:
\begin{equation}
    f_{\rm IA}(z)=a_0 +a_1z+a_2z^2 +a_3z^3 +a_4z^4.
    \label{IA_poly}
\end{equation}
Further details regarding this choice can be found in \autoref{app:IA}. 

Finally we consider the cross-correlation between GCph and WL, also referred to as
 galaxy-galaxy lensing, with the following  angular power spectrum:
\begin{equation}
    C^{\rm{GCph}\times\rm{WL}}_{ij}(\ell)=C^{\rm{GCph}\times\gamma}_{ij}(\ell)+C^{\rm{GCph}\times\rm{IA}}_{ij}(\ell),
    \label{eq:Cls_GC_WL}
\end{equation}
where the different terms have already been defined in Eq.~\eqref{window_gal}, Eq.~\eqref{window_shear} and Eq.~\eqref{window_IA}.

\subsection{Gravitational Waves observables}\label{subsec:gw_observables}

In the  GW sector we consider the weak lensing of GWs (GW-WL) \citep{Balaudo_2023, Garoffolo_2021} as well as their number count (GW-NC) \citep{Balaudo_2024}.
The natural framework for describing both GW-NC and GW-WL is luminosity distance space (D-space) \citep{Fonseca_2023, yang2022gravitationalwavesourceclustering}. This is because both observables fundamentally depend on the measured luminosity distance: in the case GW-NC, the signal is expressed in terms of fluctuations in the observed luminosity distance, while for GW-WL the lensing effect directly perturbs the luminosity distance of the sources as they propagate through the large-scale structure. In D-space, these effects are therefore most naturally described, particularly when redshift information is not directly available. However, in order to facilitate cross-correlations with galaxy surveys, which are naturally defined in z-space, we assume that redshift information is available, either through EM counterparts or statistical association techniques \citep{Gair_2023}, the latter providing redshift information at the population level and enabling effective redshift estimates within distance bins given a sufficiently large number of events.
While this z-space treatment ensures consistency with galaxy observables and simplifies joint analyses, it is important to note that, from a theoretical standpoint, the underlying quantities are still rooted in D-space, and neglecting this can lead to differences in the resulting angular power spectra.

 These observables include several contributions that influence the final form of the angular power spectrum, both from leading and secondary perturbative effects. These can be compactly represented through the following expression: 
\begin{equation}
S_{\rm obs}^i(k) = \int^{\eta_0}_0 d\eta\, W^i(\eta)\mathcal{S}{\rm src}(\eta,k)\, j_\ell(kd_C), 
\label{eq:sources_GWs} 
\end{equation} 
where $W^i(z) = |d\eta/dz| W^i(\eta)$, $\rm obs = [GW\text{-}NC, GW\text{-}WL]$ denotes the observable considered and $\mathcal{S}_{\rm src}$ encapsulates the source window function contributing to each case. The angular power spectrum for the auto- and cross-correlations of the GW observables is then: 
\begin{equation}
    C_{ij}^{\rm A \times B} = 4 \pi \int d\,log\, k\, \langle S_{\rm obs}^i(k)S_{\rm obs}^{j*}(k)\rangle.
\end{equation}
We report the complete expression for each $\mathcal{S}_{\rm src}$ in \autoref{app:GW_sources}. 

In this work, we restrict our analysis to the dominant contributions i.e. the density contribution for the clustering and the convergence field for the lensing, in line with what we do, and is usually done,  for the galaxies \citep{Lepori_2025, Martinelli_2021} . 
We explore the impact of the remaining terms, both on the galaxy and GW side, in \autoref{app:GW_sources} using a Fisher approach, which is computationally less expensive.

 GW-WL offers a way to trace  the matter distribution in the Universe, similarly to cosmic shear. As shown in \citep{Balaudo_2023} this observable can be described via an estimator that captures the term contributed to the GW luminosity distance from the lensing convergence, which is the dominant relativistic correction in the redshift range of interest. While the use of  GW-WL in combination with WL of galaxies might seem redundant, the former is not affected by IA and has different systematics, therefore it can help breaking degeneracies among parameters and tighten constraints on the cosmological ones. 

GW-NC can be defined as the local abundance in the number of detected GW sources in the sky and has already been explored in \citep{Balaudo_2024, Fonseca_2024}. As GWs trace the underlying distribution of matter, GW-NC provides information on the clustering which is complementary to that of to  GC and affected by different systematics.
The main contribution to the source window comes from the density term, denoted as $\delta N_{\rm GW}$. As for the galaxies, the matter distribution from GW-NC is traced with a bias $b^{\rm GW}(z)$. At first approximation we assume that it aligns with the bias in the galaxy distribution, considering that GWs events occur in galaxies:
\begin{equation}
    b^{\rm GW}(z)=b^{\rm GW}_0 +b^{\rm GW}_1z+b^{\rm GW}_2z^2 +b^{\rm GW}_3z^3\,.
    \label{bias_GW}
    \end{equation}
The other contributions arise from relativistic effects that perturb GW propagation, \cdlt{which are implemented in our code, even if not included in the main analysis due to computational reasons, and summarized in \autoref{app:GW_sources}.} 
 \begin{figure}[t]
    \centering
    \includegraphics[width=0.8\textwidth]{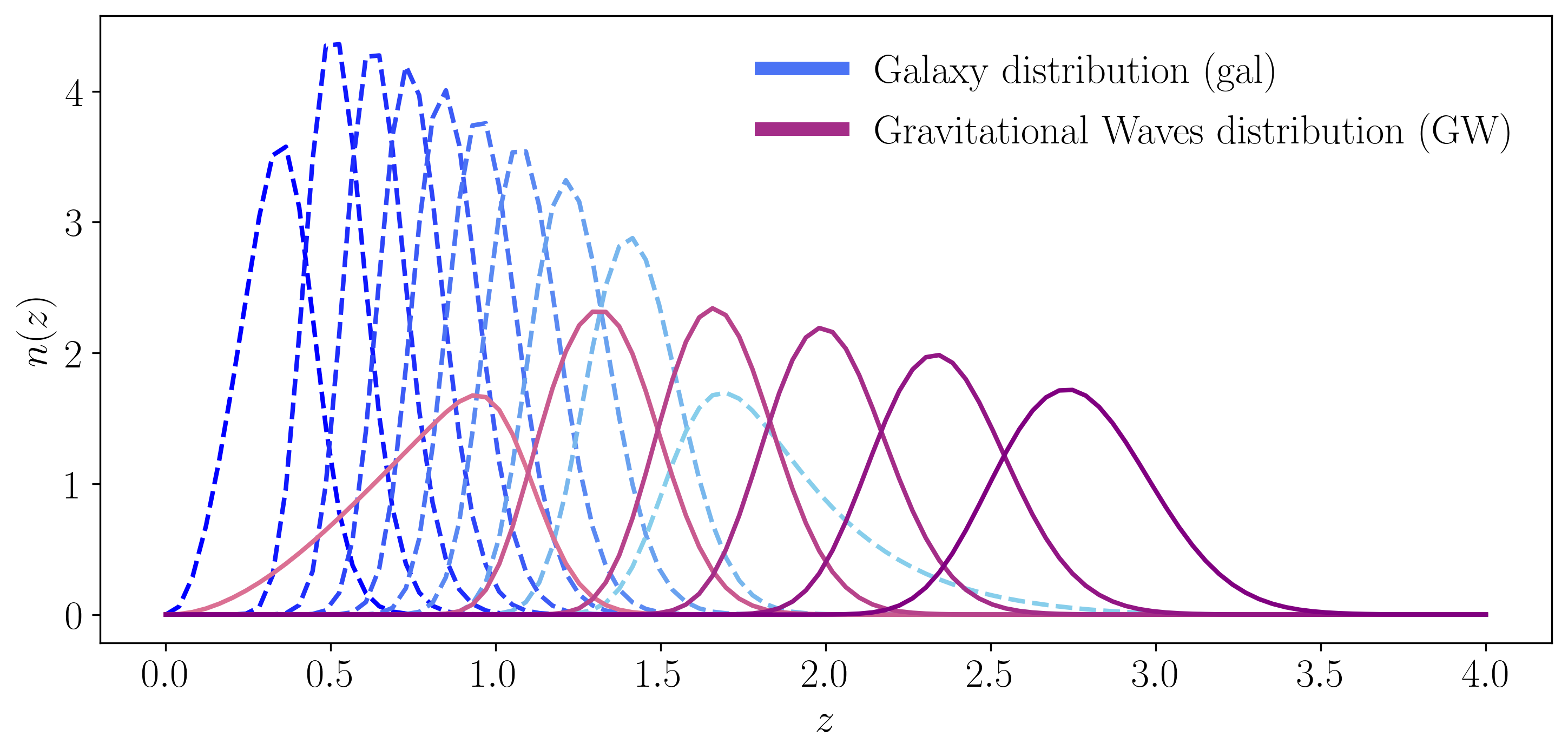}
    \caption{Representative redshift distributions of galaxies and gravitational waves for the specifics in \autoref{N_gal} and \autoref{eq:n_gw} with $z_0=\frac{0.9}{\sqrt{2}}$ and $z_0=1.5$ respectively. The galaxy one is divided in $10$ redshift bins as in \citep{EuclidPreparationForecast}, while the gravitational waves one into $6$.}
    \label{fig:src_distribution}
\end{figure}

\subsection{Cross-correlation of photometric large-scale structure surveys with gravitational waves}\label{subsec:cross_observables}

\begin{figure}[t]
    \centering
    \includegraphics[width=1.\textwidth]{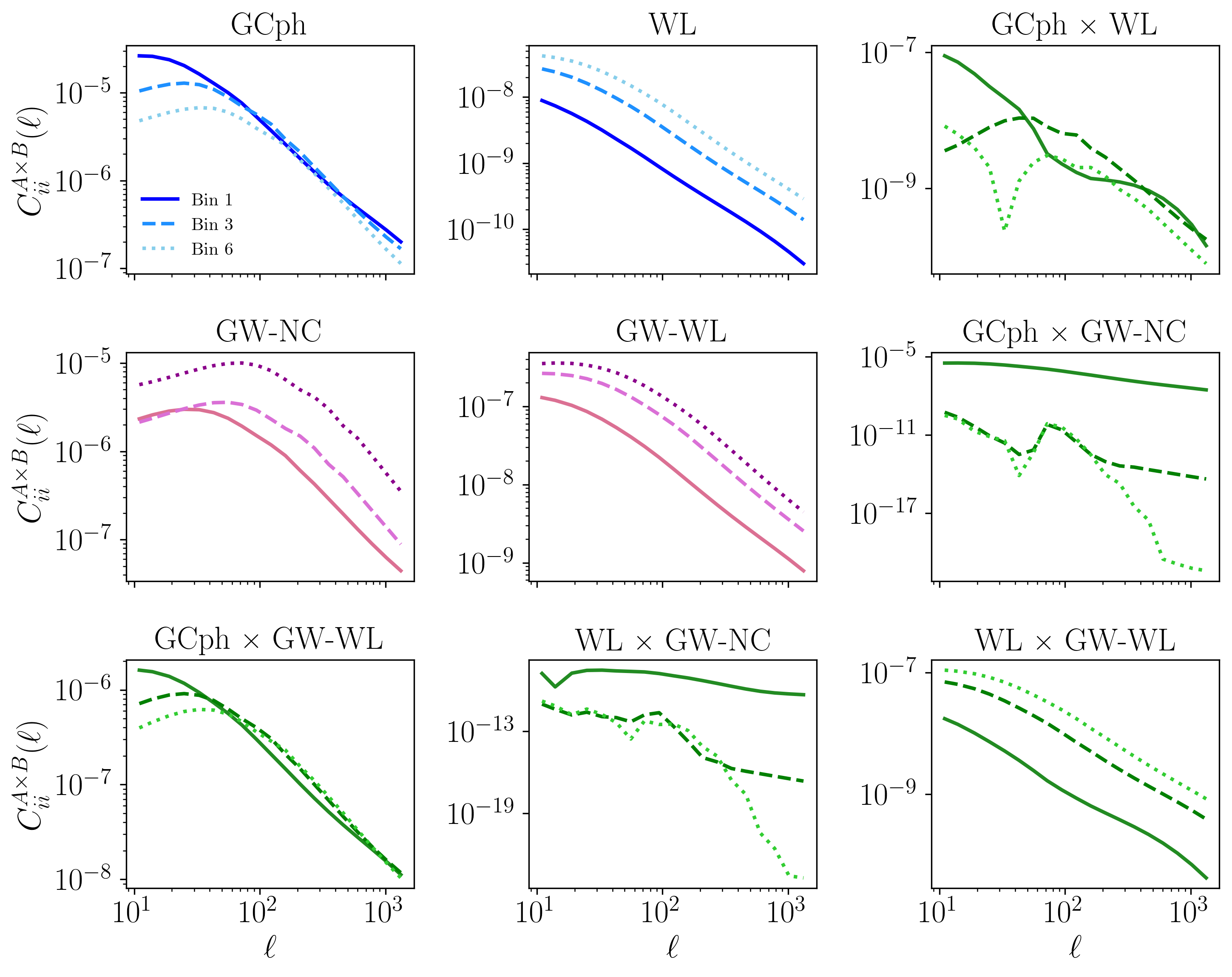}
    \caption{Contribution to the angular power spectra for LSS$\times$GW in 3 redshift bins $i=[1,3,6]$, evaluated taking the observables distribution shown above in a $\Lambda$CDM scenario. The plot displays 9 spectra corresponding to galaxy observables (blue), GW observables (purple), and their cross-correlation (green), we do not show the GW-NC$\times$GW-WL term. However, the full $10 \times 2$pt is considered in the analysis.}
    \label{fig:Cls_plot}
\end{figure}

Finally, we consider the cross-correlations between GW and galaxy observables (hereafter collectively referred to as LSS$\times$GW, or 10$\times$2pt), namely the following angular cross-spectra:
\begin{itemize}
    \item GCph $\times$ GW-WL: $C^{\rm{GCph}\times\rm{GW-WL}}_{ij}(\ell)$, that will depend on \autoref{window_gal} and the sources window functions defined in \autoref{app:GW_sources};
    \item GCph $\times$ GW-NC: $C^{\rm{GCph}\times\rm{GW-NC}}_{ij}(\ell)$, where the GW-NC sources can be found in \autoref{app:GW_sources};
    \item  WL $\times$ GW-WL: $C^{\rm{WL}\times\rm{GW-WL}}_{ij}(\ell)=C^{\rm{GW-WL}\times\gamma}_{ij}(\ell)+C^{\rm{GW-WL}\times\rm{IA}}_{ij}(\ell)$;
    \item  WL $\times$ GW-NC: $C^{\rm{WL}\times\rm{GW-NC}}_{ij}(\ell)=C^{\rm{GW-NC}\times\gamma}_{ij}(\ell)+C^{\rm{GW-NC}\times\rm{IA}}_{ij}(\ell)$;
\end{itemize}
In \autoref{fig:src_distribution}, we show an example of galaxy and GW source distributions, divided into 10 and 6 redshift bins respectively. In \autoref{fig:Cls_plot}  we display the different contributions to the angular power spectra for $\Lambda$CDM  at three representative redshift bins $i = \{1, 3, 6\}$, using the source distribution of \autoref{fig:src_distribution}. The figure highlights the complementarity between GW observables and LSS probes, showing that cross-correlations between the two can carry significant information. In particular, correlations such as GW--NC $\times$ GCph and GW-WL $\times$ WL exhibit non-negligible amplitudes over a wide range of multipoles, illustrating the potential of GW-LSS cross-correlations to enhance cosmological constraints. Some small-scale features visible in a few of the cross-spectra, in particular the presence of spikes, may be related to sign changes in the spectra, which appear as sharp features when plotted on a logarithmic scale. However, these features occur at multipoles larger than those considered in our analysis. As discussed in \autoref{sec:data_GW}, we impose a conservative scale cut at $\ell = 200$ for GW observables, ensuring that these features do not impact our results.

\begin{figure}[t]
    \centering
    \includegraphics[width=1.\linewidth]{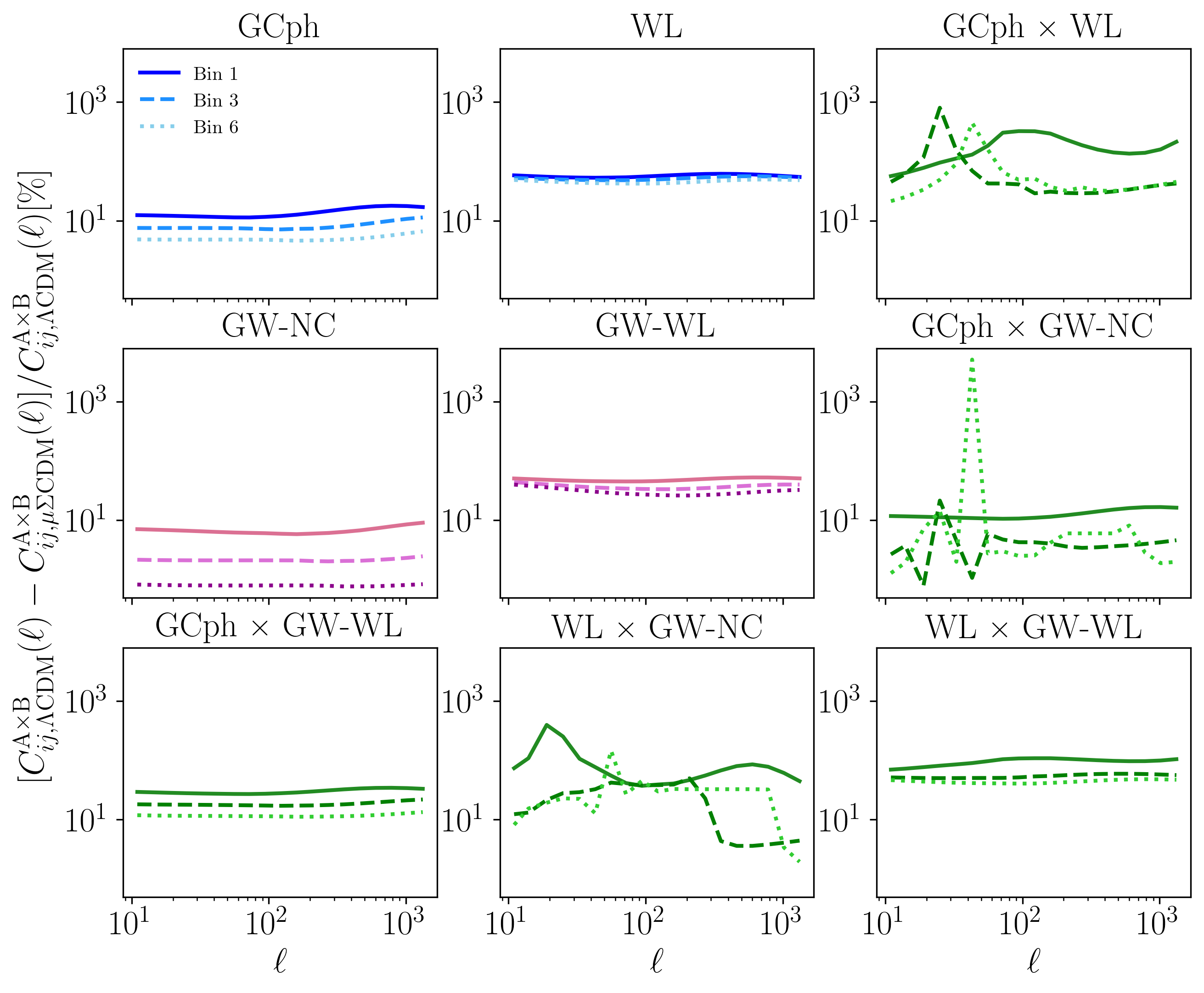}
    \caption{Impact of the cosmological model on the angular power spectra coefficients for galaxies (blue), from GWs (purple), and their cross-correlation (green). The figure shows the relative difference \cdlt{expressed in percentage units} in $C_{ij}^{\rm AB}(\ell)$, as defined in Eq.~\eqref{eq:C_ls}, between a standard $\Lambda$CDM cosmology and a $\mu\Sigma$CDM model. The source distributions used are the same as in \autoref{fig:Cls_plot}. The $\mu\Sigma$CDM model corresponds to a $\Lambda$CDM background with modified growth and lensing parameters, characterized by \cdlt{$\mu_0=0.71$ and $\Sigma_0 =0.19$. These values are the $3\sigma$ upper limits of \citep{DESy3_MG} Table IV 3rd column.}}
    \label{fig:Cls_muSigma}
\end{figure}

In \autoref{fig:Cls_muSigma} we examine the $\mu\Sigma$CDM framework, defined by a 
standard $\Lambda$CDM background and by modifications to clustering and lensing 
encoded in the functions $\mu$, defined in Eq.~\eqref{eq:mu} and $\Sigma$, defined in Eq.~\eqref{eq:Sigma}. 
The angular power spectra shown in the figure are computed for 
\cdlt{$\mu_{0}=0.71$ and $\Sigma_{0}=0.19$, values corresponding to the $3\sigma$ upper limits quoted in the 3rd column of
Table~IV in \citep{DESy3_MG}}. We consider the predictions for all auto- and cross-angular spectra in three redshift bins, and display the fractional difference with respect to the $\Lambda$CDM predictions.
Overall, we find that the fractional deviations can reach several orders of 
magnitude depending on the observable, demonstrating that angular power spectra are sensitive to deviations from the standard cosmology, highlighting the constraining power of this multi-probe analysis as test of GR. 
\cdlt{We stress that the large amplitudes arise because we intentionally assume relatively large deviations from $\Lambda$CDM, adopting values of $\mu$ and $\Sigma$ close to their current upper bounds in order to illustrate the sensitivity of the angular power spectra to departures from GR.}
Also, a few notable trends emerge from the \autoref{fig:Cls_muSigma}. The WL and GW-WL panels show almost identical fractional differences across all redshift bins. This is related to the definition of the lensing as an integrated quantity. As a consequence the contributions arise from the entire line of sight rather than from a single redshift slice. In the $\mu\Sigma$CDM framework, the modification entering through $\Sigma$ produces an almost uniform rescaling of the signal in every tomographic bin. As a result, the lensing-related observables respond in a 
nearly bin-independent way. In contrast to lensing, the clustering observables (GCph and GW-NC) display redshift-dependent fractional differences. This is expected, since these probes are sensitive to the matter density fluctuations at the specific redshift of each tomographic bin rather than to an integrated quantity. The parameter $\mu$ alters the growth of structure, so its effect varies with the redshift at which the density field is observed and with the scales probed by each angular spectrum. As a result, clustering observables respond much more strongly to departures from GR and exhibit pronounced bin-to-bin variations, making them particularly sensitive to the growth-related aspects of modified gravity. \cdlt{Also, we notice that around $\ell=40$ the relative percentage difference shows a peak in the GCph$\times$GWC spectra. After comparing the shape of the spectra in the $\Lambda$CDM scenario with the one computed in the $\mu\Sigma$CDM scenario, we notice that the overall shape is not affected while there is a difference in the amplitude in the 6-th redshift bin around $\ell=40$. We conclude that this can be addressed to a numerical issue in the integration}. As already seen in \autoref{fig:Cls_plot}, the spikes visible in some of the cross-correlations are associated with sign changes in the corresponding angular power spectra and are enhanced by the logarithmic scale; however, they occur at multipoles beyond those considered in our analysis and therefore do not affect our results.

\section{Methodology}
\label{sec:methodology}
We wish to explore the constraining power of the observables introduced in \autoref{sec:observables} in the context of upcoming galaxies and GW surveys data. To this extent, we consider two approaches:
 the Fisher matrix \citep{Fisher}, which uses theoretical predictions and specifications of the surveys to approximate the shape of the posterior around the maximum (represented by a fiducial cosmology);  full reconstruction of the posterior distribution  achieved by sampling the parameter space and fitting it to simulated data representative of the forthcoming data. In this section we describe these two techniques. 

\subsection{Evaluation of the posterior probability distribution} 
\label{sec:posterior}
While the core of our analysis is performed on synthetic data through a full sampling of the posterior, we use the Fisher matrix formalism in two instances. First, to to explore  the impact of the different settings for the GWs data (see \autoref{sec:data_GW}). Second,  to quantify the importance of the relativistic corrections in the GW-NC which we neglect in the main analysis, as described in \autoref{app:GW_sources}.

\cdlt{The Fisher matrix approach assumes a Gaussian likelihood for the harmonic coefficients $a_{\ell m}$ and is computationally efficient, providing an initial estimate of how well the parameters of a given model can be constrained and which degeneracies might arise. The elements of the Fisher matrix are:
\begin{equation}
F_{\alpha\beta}
= \frac{1}{2} \mathrm{Tr}\left[
\frac{\partial \rm{C}}{\partial \theta_\alpha}
\rm{C}^{-1}
\frac{\partial \rm{C}}{\partial \theta_\beta}
\rm{C}^{-1}
\right]
+ 
\frac{\partial \mu}{\partial \theta_\alpha}
\rm{C}^{-1}
\frac{\partial \mu}{\partial \theta_\beta}.
\end{equation}
where $\mu$ represents the mean of the data, and ${\rm C}$ is the expected covariance matrix for the data.
In the case of 3$\times$2pt analyses, the data vector is given by the set of angular power spectra $C_\ell^{ij}$, whose expectation value defines the mean of the data vector, $\mu = C_\ell^{ij}( \theta)$.
Although the covariance matrix formally depends on the cosmological parameters, this dependence is typically weak and is neglected in Fisher forecasts. Under this approximation, only the second term contribute. For the calculation of the covariance we refer to \cite{Euclidpreparation_Sciotti}.}

In general the posterior of the LSS parameters is not gaussian and it is necessary to perform a full sampling. In this work we build a complete pipeline to enable this for the 10 $\times$ 2 pt analysis. We start creating synthetic GW and LSS data representative of the upcoming and ongoing surveys, as described in detail in the next subsections.  We create the theoretical predictions with a customized version of \texttt{MGCAMB}\footnote{Soon available as a feature in the main \url{https://github.com/sfu-cosmo/MGCAMB}} \cite{Zhao_2009,Hojjati:2011ix,Zucca:2019xhg,Wang:2023tjj}, in which we implemented  the calculation of the source terms for both GW-WL and GW-NC. Further details on the validation of the code can be found in \autoref{app:validation}.
To sample the full posterior distribution, we use the nested sampler \texttt{Nautilus} \citep{nautilus}, configured with 4,000 live points and 16 trained neural networks. The pipeline is assembled through the log-likelihood evaluation model wrapper of the Bayesian framework \texttt{Cobaya}\footnote{\url{https://github.com/CobayaSampler/cobaya}} \citep{Cobaya}, for which we build an external likelihood\footnote{\url{https://github.com/chiaradeleo1/GWXLSS}, the code will be available upon publication.} able to compare LSS and GW data with theoretical predictions. The full \texttt{GW$\times$ LSS} pipeline integrates the generation of synthetic data, \cdlt{that consists of the tomographic angular power spectra for the four observables considered in this work, i.e GCph, WL, GW-NC and GW-WL. These data mimic the full tomographic reconstruction of the observables. The pipeline also allows one to compute the associated covariance matrix as well as the redshift distributions of galaxies and gravitational waves, $n^{\rm gal}(z)$ and $n_{\rm GW}(z)$.} The calculation of all auto- and cross-angular power spectra is performed through \texttt{MGCAMB} with the GWs source terms, which provides the theoretical predictions for both standard and modified-gravity scenarios. 
The external \texttt{Cobaya} likelihood module is then interfaced with the theory module to evaluate the theoretical predictions at each point in parameter space, and with \texttt{Nautilus} to sample the full posterior distribution.

\subsection{Synthetic data generation}

In this section, we detail how we generate synthetic data representative of ongoing and upcoming LSS and GW surveys.

\subsubsection{Photometric Large-Scale Structure survey}
\label{sec:data_gal}

Following what was done in \citep{Balaudo_2023}, to simulate the photometric probes, we assume the galaxy distribution to be:
\begin{equation}
    {\rm n^{gal} }(z)=\Biggl(\frac{z}{z_0}\Biggr)^2 e^{-(\frac{z}{z_0})^\frac{3}{2}}\,,
    \label{N_gal}
\end{equation}
with $z_0=\frac{0.9}{\sqrt{2}}$ \citep{Casas23}. This distribution is then divided into $10$ equipopulated redshift bins, following \citep{Casas23}, to perform the tomographic analysis. 
\cdlt{The adopted survey specifications to simulate the redshift distribution are chosen to reproduce a Euclid-like configuration, namely $f_{\rm sky}=0.35$, $N_{\rm bins}^{\rm gal}=10$, $n_{\rm gal}=30\,{\rm arcmin}^{-2}$, and $\sigma_\epsilon=0.3$, consistently with Table~4 of \cite{EuclidPreparationForecast}.}
We consider the following  error on the redshift of galaxies:
\begin{equation}
    \sigma_z^{\rm{gal}}=0.05(1+z).
    \label{galaxy redshift error}
\end{equation}
For the covariance we take into account only the Gaussian terms, neglecting the non-Gaussian components,
\cdlt{where by non-Gaussian components we refer to the additional contributions to the covariance arising from higher-order correlators of the underlying density field, as the super-sample covariance, which go beyond the Gaussian approximation. While LSS fields are not strictly Gaussian, we approximate them as Gaussian for simplicity and computational efficiency in this analysis \cite{Euclidpreparation_Sciotti}.} In this approximation the covariance can be written as:
\begin{multline}
{\rm Cov}[ {\rm C}_{ij}^{\rm AB}(\ell), {\rm C}_{ij}^{\rm A'B'}(\ell ')] = \\
 \frac{[{\rm C}_{ik}^{\rm AA'}(\ell) + \mathcal{N}_{ik}^{\rm AA'}(\ell ')] [{\rm C}_{jl}^{\rm BB'}(\ell) + \mathcal{N}_{jl}^{\rm BB'}(\ell ')] + [{\rm C}_{il}^{\rm AB'}(\ell) + \mathcal{N}_{il}^{\rm AB'}(\ell)] [{\rm C}_{jk}^{\rm BA'}(\ell) + \mathcal{N}_{jk}^{\rm BA'}(\ell ')])}{(2\ell+1)f_{sky}\delta_\ell},
\label{eq:covariance matrix}
\end{multline}
where $\rm A,A',B,B'=[GCph,WL]$, $\delta_\ell$ is the width of the multipole bin, and $N$ represents the instrument noise terms of the different probes that we assume not to be correlated between each other, i.e. $ \mathcal{N}^{\rm AB}_{ij}(\ell)=0$ if $\rm A \neq B$. We compute the Gaussian covariance with the publicly available code \texttt{Spaceborne}\footnote{\url{https://github.com/davidesciotti/Spaceborne_covg.git}} \citep{Euclidpreparation_Sciotti}.
For WL the noise term can be defined as \citep{Balaudo_2023}:
\begin{equation}
 \mathcal{N}^{\gamma}_{ij}(\ell)=\frac{\sigma_\epsilon^2}{\Bar{N}^{\rm{gal}}_i}\delta_{ij},
    \label{noise lensing}
\end{equation}
where $\rm {\Bar{N}^{\rm{gal}}_i}$ is the number of galaxies in the $i$-th redshift bin, $\sigma_\epsilon$ is the intrinsic ellipticity affecting shear measurements, i.e. the uncertainty due to the intrinsic shape of the observed galaxy, and we take $\sigma_\epsilon=0.3$ as in the reference paper.
For GC the noise is simply a Poissonian term, since our observable is the count of galaxies at given positions:
\begin{equation}
\mathcal{N}^{\rm GCph}_{ij}(\ell)=\frac{1}{{\Bar{\ N}^{\rm{gal}}_i}}\delta_{ij}.
    \label{noise clustering}
\end{equation}

\subsubsection{Gravitational Wave sources}
\label{sec:data_GW}
While the primary goal of this work is to investigate the potential of joint LSS$\times$GW analyses in constraining deviations from the standard $\Lambda$CDM model, we also aim to identify the GW data requirements needed to achieve an 
improvement in cosmological constraints relative to the galaxy-only case. In this context, it is important to recognize that employing GWs as a statistical 
cosmological probe, analogously to LSS observables, requires a sufficiently large number of events to enable a robust reconstruction of the underlying cosmological 
model.
\begin{figure}[h!]
    \centering
    \includegraphics[width=1\linewidth]{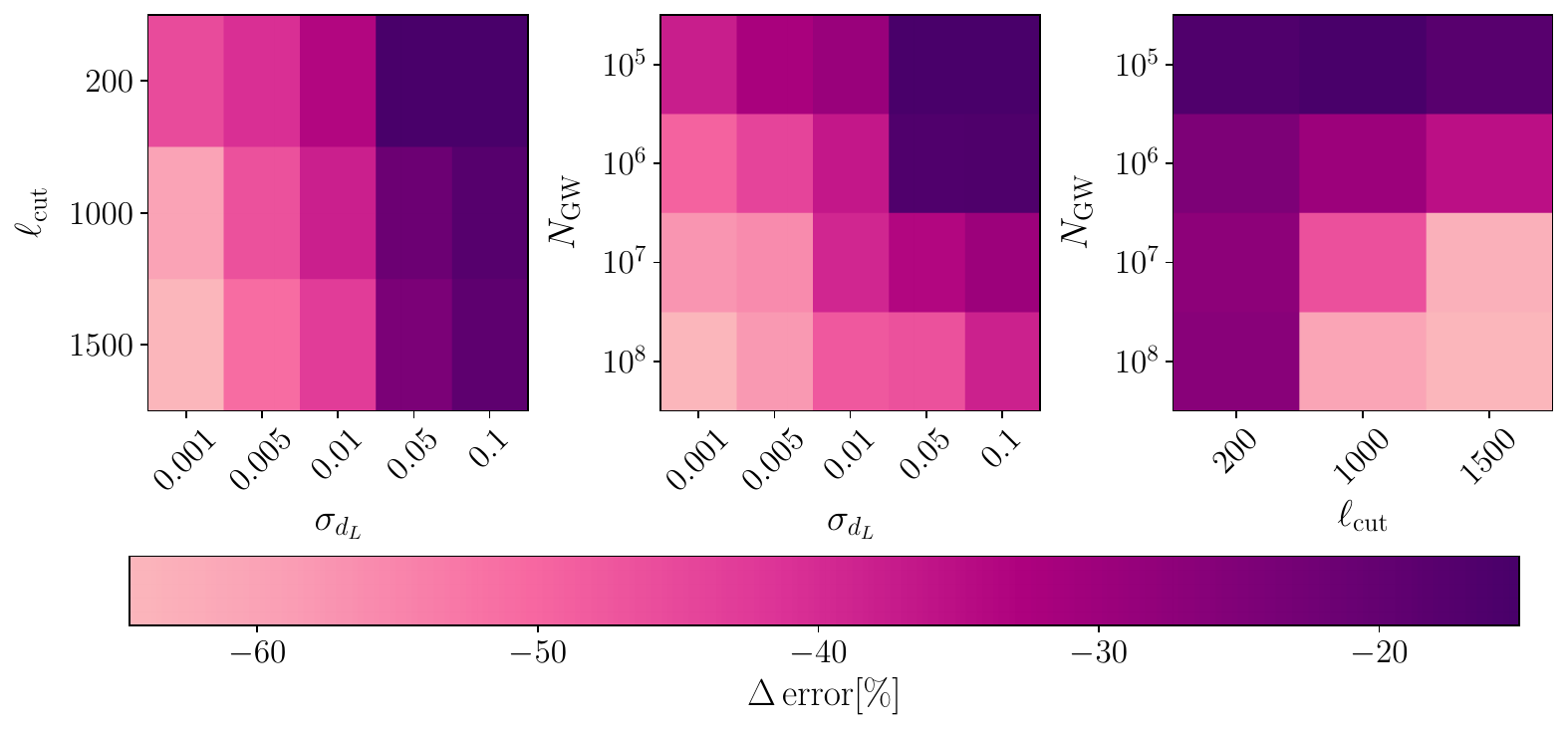}
    \caption{Relative change in the forecast error on $\Sigma_0$ when combining LSS and GW data, compared to using LSS alone. Each subplot shows the effect of varying two parameters while keeping the third fixed at the reference value used in the final configuration. The relative error is computed with Eq.~\eqref{eq:rel_err} and in the plot it is shown in percentage.}
    \label{fig:settings_GW}
\end{figure}
In \autoref{fig:settings_GW}, we illustrate the impact of different settings on the GW forecast when considering only the density term for GW-NC, and the convergence term for GW-WL, while the other relativistic effects are turned off. For each configuration, we compute the Fisher matrix and evaluate the relative error difference for a given parameter, in particular we show the case of $\Sigma_0$. This relative error, denoted as $\Delta$error, is defined as:
\begin{equation}
    \Delta \rm error = \frac{\sigma^{\rm LSS\times GW}-\sigma^{\rm LSS}}{\sigma^{\rm LSS}},
    \label{eq:rel_err}
\end{equation} where $\sigma^{\rm LSS}$ is the error estimated using LSS data alone, and $\sigma^{\rm LSS \times GW}$ is the error when combining LSS with GW data. The subplots show the impact of varying two parameters at a time, while the third is held fixed at a reference value, specifically, the value chosen for the final configuration. For each configuration we kept fixed the number of bin $N_{\rm bin}^{\rm GW}=6.$ 
As expected, the most favourable scenario corresponds to a combination of a very large number of GW events and high-precision luminosity-distance measurements. 
While such conditions may not be fully achievable in the near future, we aim to remain as realistic as possible by adopting the optimistic projections for next-generation GW detectors \citep{Abac:2025saz, Pieroni2022}. Motivated by these considerations, we adopt as our fiducial configuration 
$N_{\rm GW}=10^{6}$, $\sigma_{d_L}=1\%$, and $\ell_{\rm cut}=200$. The choice of this last parameter is related to the localization of the event, as we will explain further in this section. With this configuration we would get an improvement on the error of around the $20\%$. The impact of the inclusion of the relativistic effects is shown in \autoref{app:GW_sources}.

We consider a distribution of GW events that follows \cite{Balaudo_2023}:
\begin{equation}
    {\rm n^{GW} }(z)\propto \Biggl(\frac{z}{z_0}\Biggr)^2 e^{-(\frac{z}{z_0})^{3/2}},
    \label{eq:n_gw}
\end{equation}
that is the same expression that we expect from galaxy from Eq.~\eqref{N_gal} with $z_0$ being the survey-dependent parameter. In particular for this work we choose to take $z_0=1.5$ in order to obtain the phenomenological distribution of GWs expected from the optimistic projection of Einstein Telescope+CE \citep{Pieroni2022}. 
In the upper panel of  \autoref{fig:Cls_plot} the two distributions chosen are shown.

We assume the same expression for the covariance as in Eq.~\eqref{eq:covariance matrix} and we define the noise associated to the GW-WL as
\begin{equation}
\mathcal{N}^{\rm{GW-WL}}_{ij}(\ell)=\frac{1}{\Bar{N}^{\rm{GW}}_i}\frac{\sigma_{d_L}^2}{d_L^2}e^{\frac{\ell^2 \theta_{\rm min}^2}{2ln8}}\delta_{ij}.
    \label{noise gw}
\end{equation}
\cdlt{ As discussed in \citep{Namikawa2016}, this form is obtained under the assumption that the dominant contribution to the noise budget comes from the uncertainty in the luminosity distance, in which case the effective noise term can be approximated as a shot-noise-like contribution. There are other contributions that given the typical values of $\sigma_{d_L}$ are negligible; in the ideal case of a very low $\sigma_{d_L}$ these contributions would need to be accounted for.}
For galaxy counts, the noise associated to GW-NC can be written as
\begin{equation}
\mathcal{N}^{\rm{GW-NC}}_{ij}(\ell)=\frac{1}{\Bar{N}^{\rm{GW}}_i}e^{\frac{\ell^2 \theta_{\rm min}^2}{2ln8}}\delta_{ij}.
    \label{noise gwclustering}
\end{equation}

With respect to the galaxy case, both noise terms have an extra factor, the \textit{beam term}  $\theta_{\rm min}$,  representing the expected resolution of the sky localization of the event. From this, one can correspondingly determine the maximum  multipole to be used in the analysis of the power spectra, with $\ell \sim 180/\sqrt{\pi \Omega}$, where $\Omega$ is the area of localization related to the angular resolution through $\Omega\simeq\pi\theta_{\rm min}^2$, which implies an effective optimistic multipole cutoff  $\ell_{\rm cut}\sim\pi/\theta_{\rm min}$.
 We set $\theta_{\rm min}=30$ arcmin that roughly corresponds to $\ell_{\rm cut}\sim 200$ for both GW-NC and GW-WL. 

\subsection{Data, models and sampled parameters} \label{sec:models}
We consider two cosmological scenarios: $\Lambda$CDM and $\mu\Sigma$CDM. In \autoref{tab:fiducial_model} we summarize the fiducial values and prior range for the sampled cosmological, model and nuisance parameters for the two cases. Both models share the $5$ standard cosmological parameters, i.e. the Hubble constant $H_0$, the baryon and cold dark matter energy density $\Omega_bh^2$ and $\Omega_ch^2$, the spectral index and amplitude of the primordial power spectrum $n_s$ and $A_s$. For these we assume the fiducial value of Planck18 \citep{Planck18}. In the $\mu\Sigma$CDM case we have two extra, model parameters, $\mu_0$ and $\Sigma_0$, from Eq.~\eqref{eq:mu} and Eq.~\eqref{eq:Sigma}. The choice of prior ranges is informed by a preliminary Fisher analysis: specifically, we compute the Fisher matrix for the $\mu\Sigma$CDM case with LSS$\times$GWs and use \texttt{getdist} \citep{getdist} to obtain sampled chains from it. For parameters with flat priors, the prior interval is defined as the fiducial value $\pm 5\sigma$. For parameters with Gaussian priors, we adopt a normal distribution centered on the fiducial value with the corresponding $\sigma$ as width, with the only exception being $\Omega_bh^2$ which has a normal distribution that incorporates external information form the Big Bang Nucleosynthesis (BBN) \citep{BBN}.

In both cases, the cosmological parameters are supplemented by nuisance parameters, which account for intrinsic alignments (IA) and the biases in the galaxy and GW distributions, which we assume to be identical.
Altogether, the full analysis LSS$\times$GW   includes 14 parameters in the $\Lambda$CDM case and 16 in the $\mu\Sigma$CDM case.
\begin{table*}[htp!]
\caption{Reference values and prior distributions for the cosmological and nuisance parameters for the models listed in \autoref{sec:models}.}
\centering
\renewcommand{\arraystretch}{1.3} 
\begin{tabularx}{\textwidth}{>{\raggedright\arraybackslash}m{6cm} >{\centering\arraybackslash}m{2.5cm} >{\centering\arraybackslash}m{2cm} >{\centering\arraybackslash}m{3cm}}
\multicolumn{2}{l}{\textbf{Parameters}} & \textbf{Fiducial Value} & \textbf{Prior} \\
\midrule
\multicolumn{4}{c}{\textbf{Cosmological Parameters}} \\
\midrule
Present day Hubble constant & $H_0$ & $67.0$ & $\mathcal{U}(62.7,\,80.0)$ \\
Present-day physical baryon density & $\Omega_b\,h^2$ & $0.02244$ & $\mathcal{N}(0.0244,\,0.00031)$ \\
Present-day physical cold dark matter density & $\Omega_c\,h^2$ & $0.12$ & $\mathcal{U}(0.1031,\,0.1381)$ \\
Slope of primordial curvature power spectrum & $n_s$ & $0.96$ & $\mathcal{U}(0.9095,\,1.013)$ \\
Amplitude of primordial curvature power spectrum & $\ln(10^{10}\,A_s)$ & $3.05685$ & $\mathcal{U}(2.895,\,3.291)$ \\
Deviation of matter clustering from GR & $\mu_0$ & $0.0$ & $\mathcal{U}(-0.4,\,0.4)$ \\
Deviation of gravitational lensing from GR & $\Sigma_0$ & $0.0$ & $\mathcal{U}(-0.36,\,0.24)$ \\
\midrule
\multicolumn{4}{c}{\textbf{Photometric Sample}} \\
\midrule
Coefficients for intrinsic alignments & $10^{3}\,a_{\, i=0\ldots4}$ & \{--7.589, 2.008, --4.127, 2.918, --0.6784\} & $\mathcal{N}(-7.589,\,-0.15)$, $\mathcal{N}(2.008,\,0.04)$, $\mathcal{N}(-4.127,\,0.08)$, $\mathcal{N}(2.918,\,0.06)$, $\mathcal{N}(-0.6784,\,0.013)$ \\
Coefficients for clustering bias & $b^{\mathrm{gal}}_{\,i=0\ldots3}$ & \{0.8307, 1.1905, --0.9284, 0.4233\}&$\mathcal{U}(0.8104,\, 0.8507)$, $\mathcal{U}(1.1413,\,1.2397)$, $\mathcal{U}(-0.9827,\,-0.8740)$, $\mathcal{U}(0.4065,\,0.4401)$  \\
\midrule
\multicolumn{4}{c}{\textbf{Gravitational Wave Sample}} \\
\midrule
Coefficients for clustering bias & $b^{\mathrm{gw}}_{\,i=0\ldots3}$ & \{0.8307, 1.1905, --0.9284, 0.4233\} &$\mathcal{U}(0.8136,\, 0.8476)$, $\mathcal{U}(1.1486,\,1.2306)$, $\mathcal{U}(-0.9714,\,-0.8834)$, $\mathcal{U}(0.4095,\,0.4365)$ \\
\bottomrule
\bottomrule
\end{tabularx}
\justifying
\footnotesize{Fiducial values used to compute the synthetic data discussed in \autoref{sec:data_gal} and \autoref{sec:data_GW}. Photometric nuisance parameters use polynomial fitting formulas for galaxy bias $b^{\mathrm{gal}}$ and GW bias $b^{\mathrm{gw}}$ with coefficients $i = 0\ldots3$, and IA $a$ with $i = 0\ldots4$. Sampled parameters follow either uniform ($\mathcal{U}$) or Gaussian ($\mathcal{N}$) priors.}
\label{tab:fiducial_model}
\end{table*}

\section{Results}\label{sec:results}
In this section we discuss the  results obtained for the different models discussed in \autoref{sec:models}. 
Since this is a forecast study, our goal is to remain as realistic as possible regarding the data that would be available in an actual observation. In practice, one would work with a single, combined dataset containing all LSS and GW observables. To reflect this, we construct a complete mock LSS$\times$GW dataset generated with $\mu=0$ and $\Sigma=0$, and including multipoles up to $\ell_{\max}=1500$. When analyzing the LSS-only case (3 $\times$2 pt), we keep only the LSS auto(cross)-correlations by masking out the GW auto and cross-correlation terms. \cdlt{Since all cases analyzed in this work are constructed from a single mock dataset and its corresponding covariance matrix, this procedure guarantees internal consistency: whenever a subset of observables is considered, the non-considered auto- and cross-spectra are consistently removed from both the data vector and the covariance matrix.}
\cdlt{Since our current implementation does not allow different $\ell_{\max}$ values for different probes, the full mock dataset is generated using the galaxy $\ell_{\max}=1500$, which is higher than the one expected for GWs; therefore, we consistently remove all GW auto- and cross-spectra evaluated at multipoles $\ell>200$ from both the data vector and the covariance matrix, ensuring that the analysis properly reflects the expected signal-to-noise limitations of the GW observations at small angular scales.}
All runs are performed including observational systematics, bias and IA, while keeping only the dominant physical contributions to the signal. The impact of the relativistic effects, which are subdominant in this regime, is instead explored separately through a Fisher analysis (see \autoref{app:GW_sources}).

\subsection{$\Lambda$CDM} \label{sec:LCDM_res}
\begin{figure}[h!]
    \centering
    \includegraphics[width=1\linewidth]{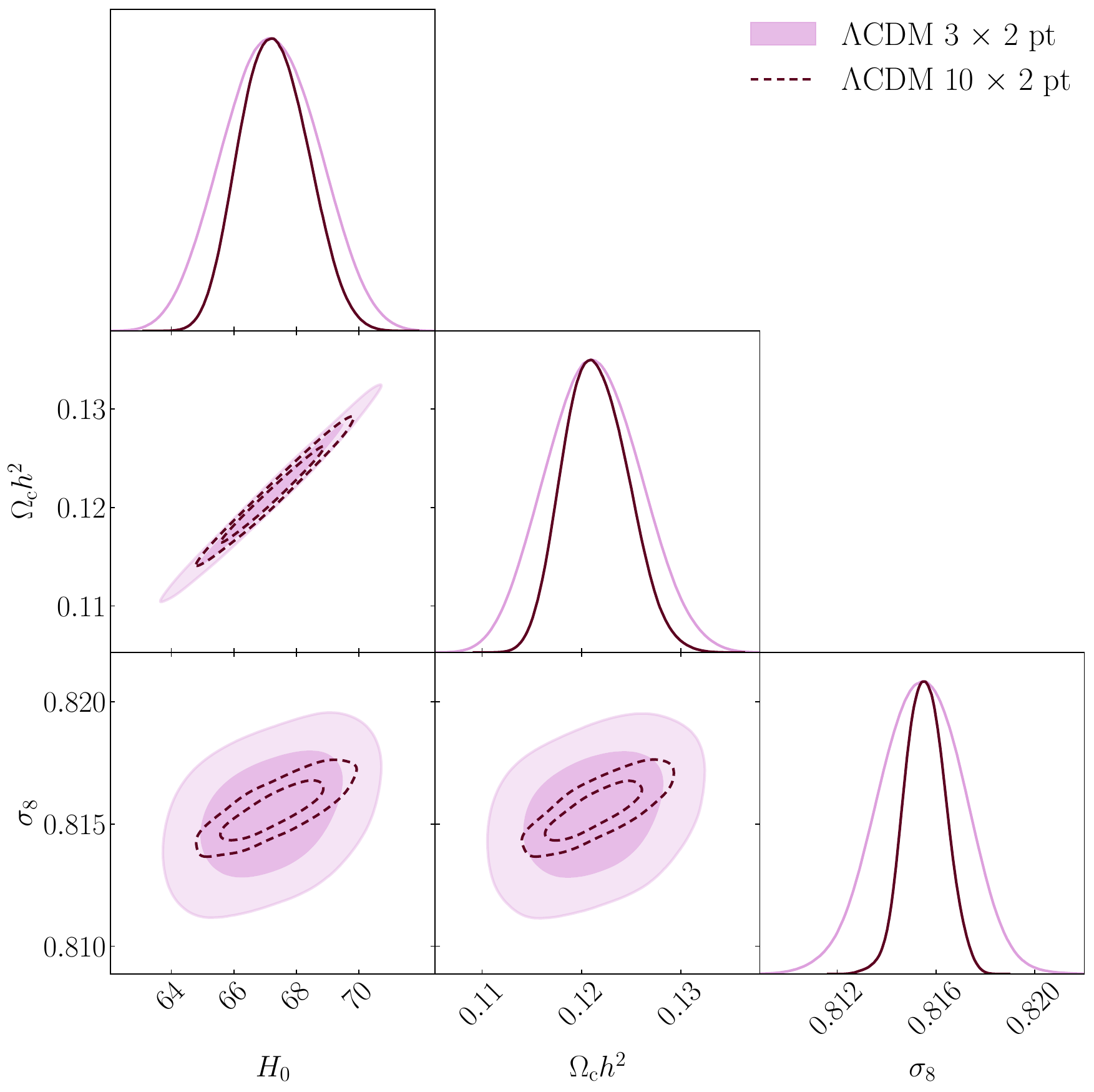}
    \caption{Forecast of the constraints for the $\Lambda$CDM model for the cosmological parameters $H_0,\,\Omega_ch^2,\,\sigma_8$ using LSS only -3 $\times$ 2 pt- (pink shaded) and LSS and GW probes -10 $\times$ 2 pt- (bordeaux-dashed).}
    \label{fig:LCDM_cosmo}
\end{figure}
 
In \autoref{tab:Results} we report the $68\%$ confidence intervals for all parameters in the $\Lambda$CDM scenario. We find the constraints obtained from the LSS-only configuration (3 $\times$2 pt) to be consistent with past \cite{Blanchard-EP7} and latest \cite{GCH24}  Euclid forecasts, despite differences in the analysis choices such as the number of redshift bins, the adopted scale cuts, and the treatment of systematics.
\noindent

In addition to serving as our baseline cosmological model, $\Lambda$CDM  plays a key role in validating the statistical pipeline, by providing a direct check that our likelihood implementation, sampling procedure, and theoretical predictions reproduce results in line with those in the literature. This guarantees that the improvements observed in the modified-gravity analysis genuinely arise from the inclusion of GW probes, rather than from any
artefact of the statistical machinery. In this sense, $\Lambda$CDM acts both as a benchmark for comparison and as an internal consistency test for the full analysis pipeline.

The marginalized posteriors of the cosmological parameters ($H_0$, $\Omega_c h^2$, $\sigma_8$)  are shown in \autoref{fig:LCDM_cosmo}, where $\sigma_8$ is a derived parameter obtained from \texttt{CAMB}.
As shown in the figure, the inclusion of GW produces tighter cosmological constraints, despite the relatively large parameter space induced by the set of nuisance parameters. In particular this can be seen in $\sigma_8$, one of the primary parameters characterizing LSS. This improvement is primarily driven by the ability of GW data to break degeneracies between cosmological and nuisance parameters, most notably the galaxy bias parameters. This behaviour was already explored in \citep{Canas-Herrera:2019npr, Canas-Herrera:2021qxs}. The origin of this improvement becomes clear when inspecting the 2D posteriors involving the nuisance parameters, most notably $b_0$ and $a_0$, as shown, respectively, in the right and left panels of \autoref{fig:sigma8_nuis_all} in pink. In both cases, we observe a strong degeneracy between the clustering bias amplitude $b_0$ (for GCph), the first-order IA polynomial expansion parameter $a_0$ (for WL), and the amplitude of matter fluctuations $\sigma_8$. Once GW observables are included, this degeneracy is partially broken. This behaviour is expected. GW-NC probes the same underlying matter distribution as GCph, but it is affected by different observational systematics and, crucially, by a different bias relation. It therefore acts as an independent tracer of the LSS, helping to disentangle cosmology from galaxy-specific nuisance parameters. 

A similar argument applies to GW-WL: unlike galaxy WL, the GW lensing signal is completely unaffected by IA, providing clean information on the projected mass distribution and further contributing to breaking the $a_0$–$\sigma_8$ degeneracy. We emphasise that $b_0$, as the constant term of the polynomial galaxy-bias model, determines the overall amplitude of the galaxy density contrast, and $a_0$ plays an analogous role for the  IA contribution. Both parameters therefore carry an intrinsic degeneracy with the fluctuation amplitude $\sigma_8$, making this degeneracy particularly sensitive to improvement when independent tracers such as GW-NC and GW-WL are added. This interpretation is quantitatively supported by the correlation coefficients extracted with \texttt{getdist}. In the LSS–only case, the correlation between $a_0$ and $\sigma_8$ is ${\sim}0.56$, but it is reduced to ${\sim}0.31$ once GW information is added. A similar trend is observed for the clustering sector: the $b_0$–$\sigma_8$ correlation decreases from ${-}0.78$ in the LSS–only analysis to ${-}0.53$ when including GWs. These reductions confirm that GW probes indeed help break the dominant amplitude degeneracies that limit the constraining power of photometric LSS observables alone.


\subsection{$\mu\Sigma$CDM} \label{sec:muSigma_res}

\begin{figure}[t]
    \centering
    \includegraphics[width=1\linewidth]{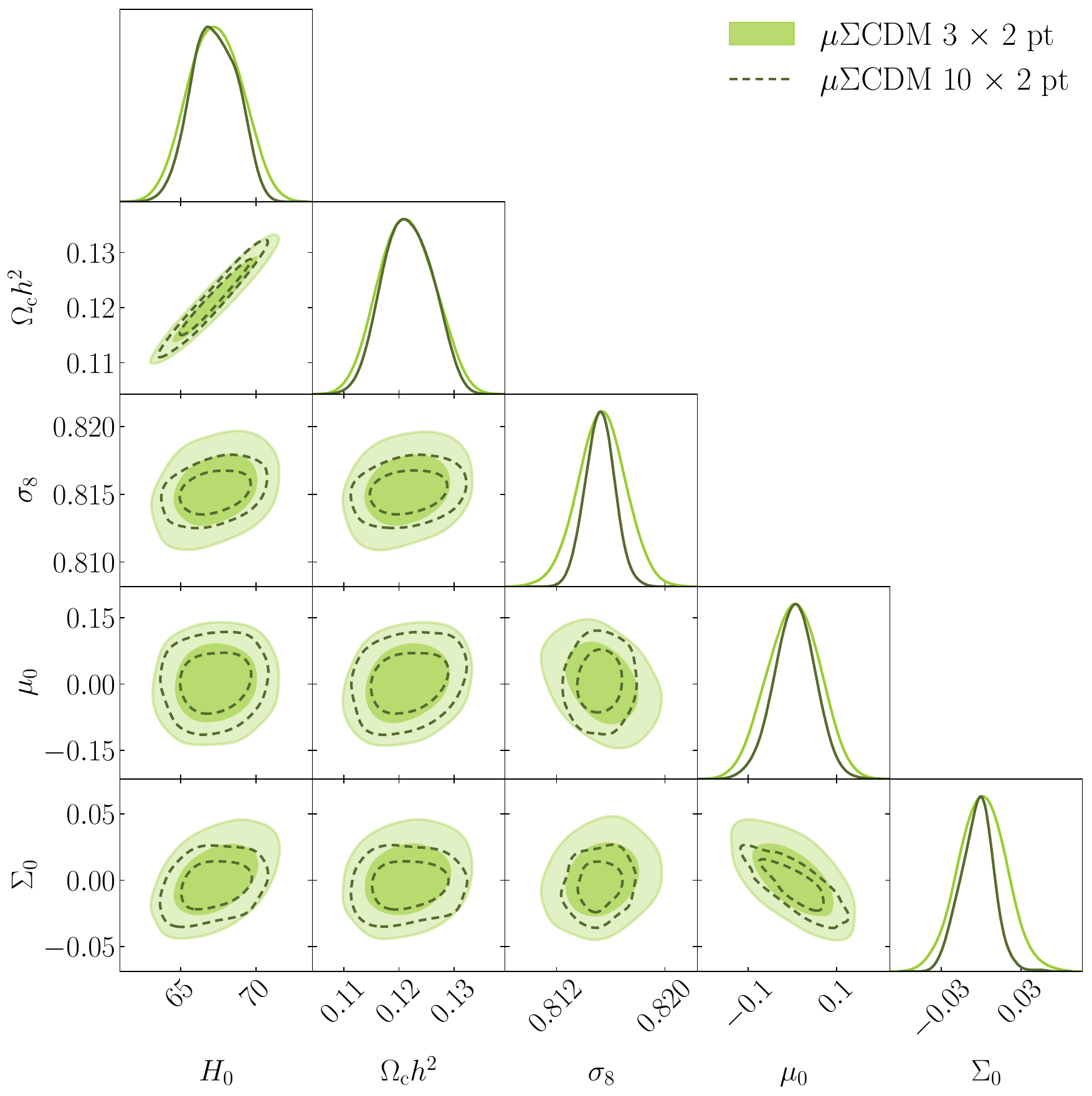}
    \caption{Forecast of the constraints for the $\mu\Sigma$CDM model for the cosmological parameters $H_0,\,\Omega_ch^2,\,\sigma_8,\,\mu_0,\, \Sigma_0$ using LSS only -3 $\times$ 2 pt- (yellowgreen shaded) and LSS and GW probes -10 $\times$ 2 pt- (dark green-dashed).}
    \label{fig:muSigma_cosmo}
\end{figure}

\begin{figure}[h!]
    \centering
    \includegraphics[width=1\linewidth]{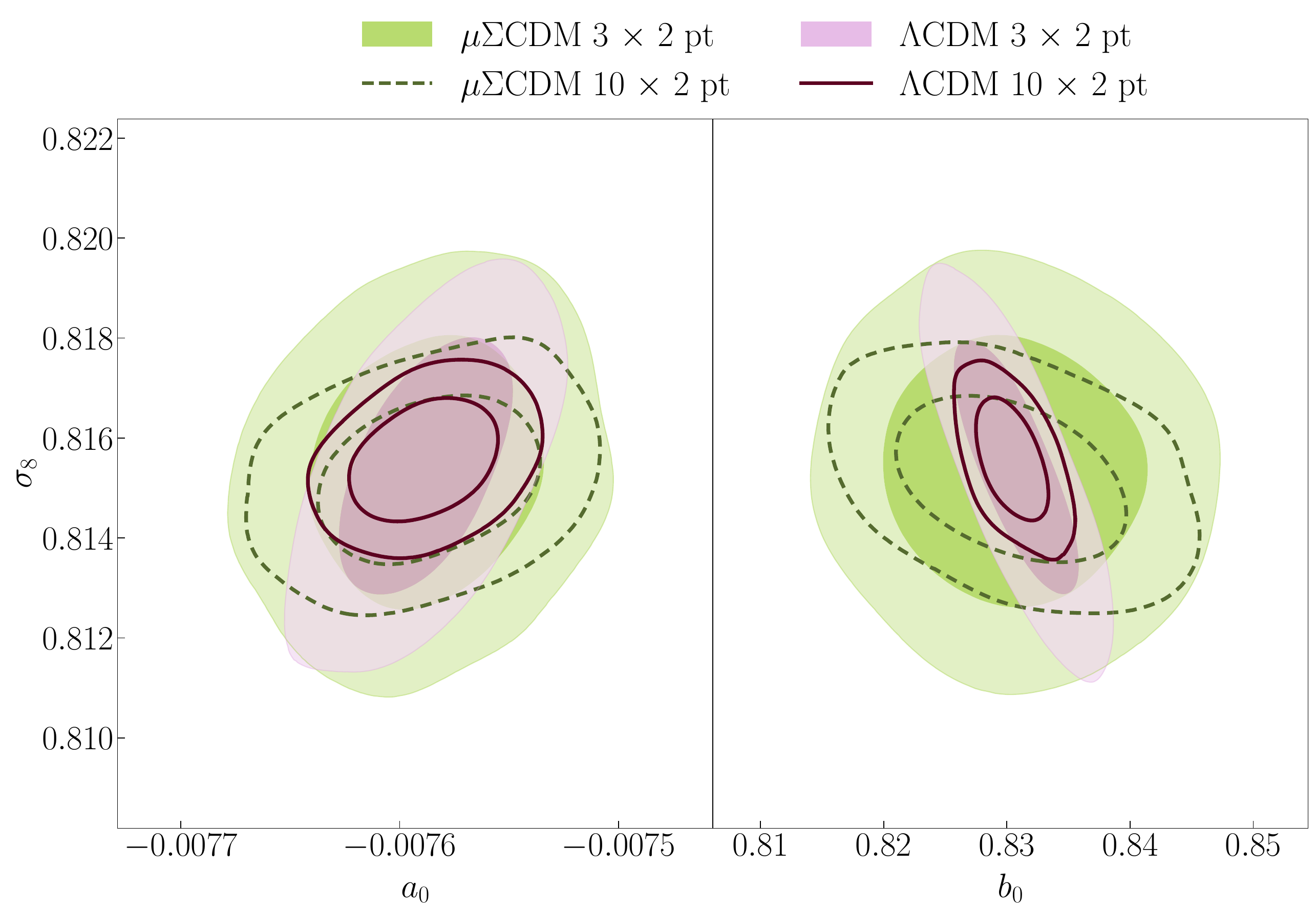}
    \caption{Two-dimensional posterior distributions for the cosmological parameter $\sigma_{8}$ 
and the IA first-order amplitude $a_{0}$ (left panel), and for $\sigma_{8}$ and the galaxy-bias amplitude $b_{0}$ (right panel). We show the constraints obtained for the $\Lambda$CDM model using LSS data only -$\Lambda$CDM 3 $\times$2 pt- (pink solid contours) and the joint LSS+GW analysis ($\Lambda$CDM 10 $\times$2 pt- (bordeaux dashed contours), and for the $\mu\Sigma$CDM case with galaxy only -$\mu\Sigma$CDM 3 $\times$ 2 pt- (yellowgreen shaded) and in combination with GWs -$\mu\Sigma$CDM 10 $\times$ 2 pt- (dark green-dashed).
}
    \label{fig:sigma8_nuis_all}
\end{figure}

\begin{figure}[h!]
    \centering
    \includegraphics[width=1\linewidth]{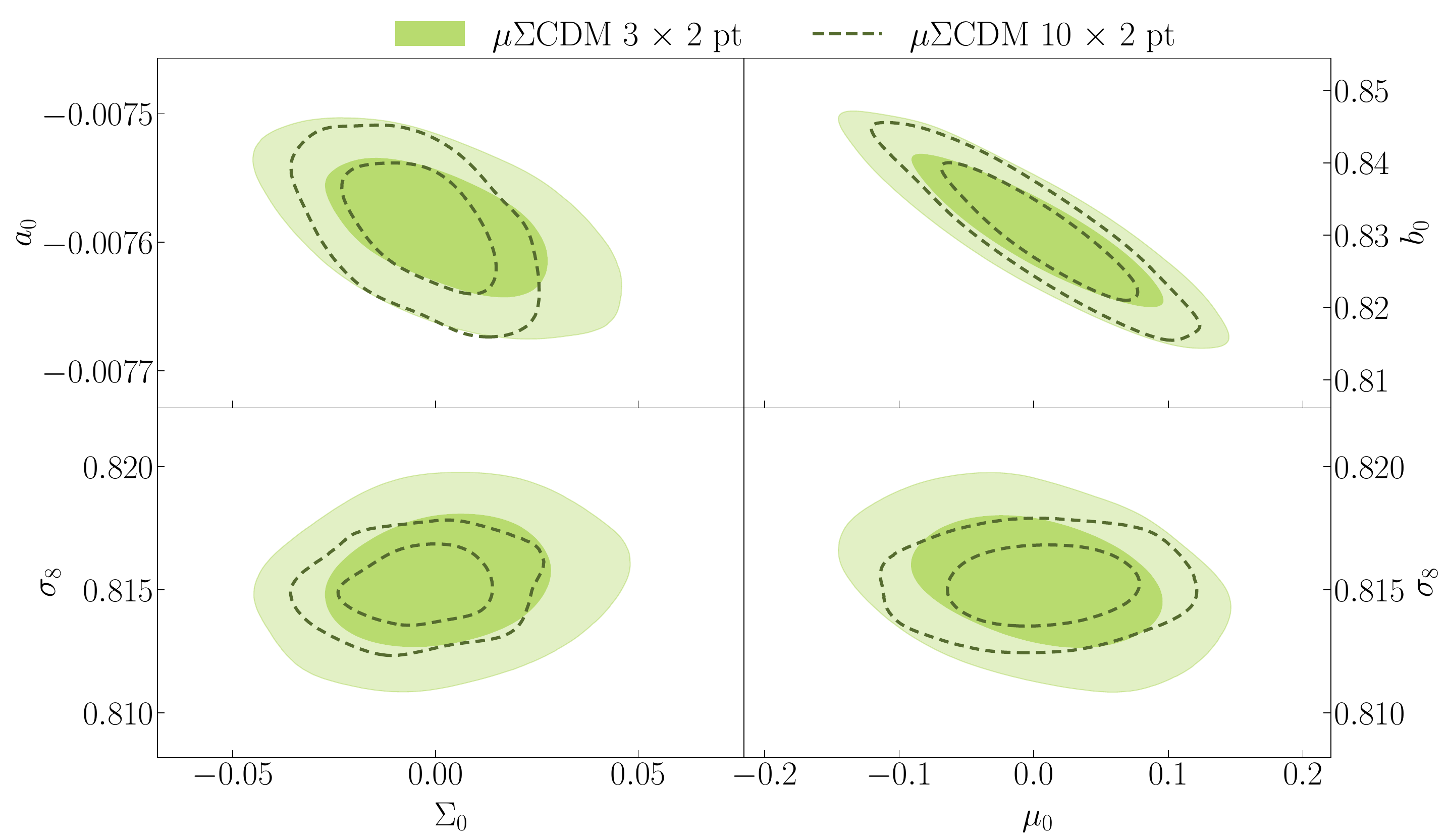}
    \caption{Two-dimensional posterior distributions for the cosmological parameter $\Sigma_0$ and the main contribution to the IA, $a_0$, (upper left panel) , $\sigma_8$ and $a_0$ (lower left panel),  $\mu_0$ and the main contribution to the galaxy bias $b_0$ (upper right panel) and $\sigma_8$ and $b_0$ (lower right panel). We show the results for the $\mu\Sigma$CDM cosmological model using LSS only -3 $\times$ 2 pt- (yellowgreen shaded) and LSS and GW probes -10 $\times$ 2 pt- (dark green-dashed).}
    \label{fig:muSigma_nuis}
\end{figure}


We now turn to the second scenario, i.e.  $\mu\Sigma$CDM. In addition to the standard $\Lambda$CDM parameters, this model includes the two modified-gravity parameters $\mu_0$ and $\Sigma_0$, introduced in Eqs.~\eqref{eq:mu} and \eqref{eq:Sigma}, which describe possible deviations from General Relativity. In total, we sample over all 16 parameters listed in \autoref{tab:fiducial_model}. As in the previous section, we compare results obtained using LSS probes only (3$\times$2 pt) with those including GW cross-correlations (10 $\times$2 pt). The resulting $68\%$ confidence intervals for all sampled parameters are similarly reported in \autoref{tab:Results}.

In \autoref{fig:muSigma_cosmo} we show the marginalized constraints on the parameters $H_0$, $\Omega_c h^2$, $\sigma_8$, $\mu_0$, and $\Sigma_0$. Compared to the $\Lambda$CDM, the improvement on the constraints with the inclusion of GW observables is milder. This is expected: on one hand, GWs provide additional cosmological information with distinct systematics; on the other hand, allowing deviations from GR introduces extra degrees of freedom that naturally broaden the parameter space. 
The difference between the two models becomes even clearer when looking at \autoref{fig:sigma8_nuis_all}. In the $\Lambda$CDM case (pink contours), the inclusion of GW information not only tightens the constraints but also rotates the $\sigma_8$-$a_0$ and $\sigma_8$-$b_0$ degeneracy directions. This change in orientation signals that GW-NC and GW-WL provide genuinely independent information on the amplitude of the LSS, enabling an efficient breaking of the degeneracy between $\sigma_8$ and the nuisance parameters. In the $\mu\Sigma$CDM model (green contours) however, no such rotation is observed: the degeneracy directions remain essentially unchanged when GWs are included in the study. This confirms the picture emerging in the upper panel of \autoref{fig:muSigma_nuis}: the additional modified-gravity parameters $\mu_0$ and $\Sigma_0$ introduce new amplitude-like degrees of freedom that mimic the observable effects of $a_0$, $b_0$, and $\sigma_8$. As a result, the parameter space becomes intrinsically more degenerate, and the GW 
probes are no longer able to disentangle $\sigma_8$ from the nuisance sector as effectively as in the $\Lambda$CDM case.

This behaviour is clearly reflected in the correlation coefficients of the parameters. 
In the LSS-only case, the MG parameters already introduce strong amplitude-like degeneracies: $\mu_0$ is very tightly correlated with the galaxy-bias 
amplitude $b_0$ (with a correlation coefficient of about $-0.88$), while $\Sigma_0$ displays a similar but less strong anti-correlation with the IA amplitude parameter $a_0$ (around $-0.55$). Both parameters also retain non-negligible 
correlations with the cosmology, in particular with $\sigma_8$, for example 
$\mu_0$--$\sigma_8 \simeq -0.32$ and $\Sigma_0$--$\sigma_8 \simeq 0.19$, indicating that they partially replicate the observable effect of changing $\sigma_8$ itself.
When GW information is included, these degeneracies are not reduced as in the $\Lambda$CDM case, instead, some of them persist or even increase.

In the lower panel of \autoref{fig:muSigma_nuis} we show the contour plots for  $\sigma_8,\mu_0$ and $\sigma_8,\Sigma_0$. It can be seen that the addition of GW data leads to a noticeable tightening of the constraint on $\Sigma_0$, only a marginal tightening of the constraint on $\mu_0$ but an interesting breaking of the degeneracy between $\sigma_8$ and $\mu_0$. However, the degeneracy between $\mu_0$ and $b_0$ remains largely intact. This behaviour is fully consistent with the  data we consider. Our GW 
catalog contains $\mathcal{O}(10^6)$ events,  which is orders of magnitude smaller than the galaxy sample contributing 
to GCph. Moreover, the GW sobservables are only used up to $\ell_{\max}=200$, a significantly more restrictive scale cut compared to the $\ell_{\max}=1500$ adopted for GCph. As a result, GW–NC probes far fewer modes, both because of the limited number of sources and because much of the small-scale information is removed by the $\ell$-cut. As such, it does not add significant constraining power on $\mu_0$, but it does help in breaking the degeneracy with $\sigma_8$ as can be seen in the lower left panel of \autoref{fig:muSigma_nuis}.
The inclusion of GW-WL introduces an observable that probes $\Sigma_0$ without the systematics of IA; therefore, even with the very conservative scale cut, GW-WL produces a clear improvement in the $\sigma_8$–$\Sigma_0$ contour, as can be seen in the lower right panel of \autoref{fig:muSigma_nuis}. 


Overall, in our setup the statistical power of GW–NC is limited. It is important to stress, however, that this limitation is strongly driven by the choice of scale cut. In our forecast, GW observables are used 
only up to $\ell_{\max}=200$, whereas most LSS forecasts rely on much more optimistic 
$\ell$–ranges, sometimes extending well beyond the regime where modelling is fully robust in particular for what concerns low redshifts \citep{euclidcollaboration2025euclidpreparationconstrainingparameterised}. 
A more careful modelling of GCph, including scale cuts as a function of redshift, nonlinear bias, and related systematics, may reduce the overall significance of this probe, thereby enhancing the relative impact of GW–NC. Furthermore, in this work we restrict the GW-NC observable to the leading density contribution, neglecting relativistic projection effects. While this choice is adequate for a first assessment, relativistic corrections are known to introduce additional sensitivity to modified-gravity effects and could significantly enhance the constraining power of GW-NC. In addition, 
performing the GW-NC analysis directly in luminosity-distance space would introduce further dependence on the underlying gravity theory, potentially providing an 
additional handle on deviations from General Relativity.
\begin{table*}
\centering
\hspace*{-1.5cm}
\renewcommand{\arraystretch}{1.2}
\setlength{\tabcolsep}{6pt}

\begin{tabular}{l|c|c|c|c} 
\hline
\multirow{2}{*}{} 
& \multicolumn{2}{c|}{$\Lambda$CDM}   
& \multicolumn{2}{c}{$\mu\Sigma$CDM} \\
\cline{2-5}
& 3$\times$2pt & 10$\times$2pt  & 3$\times$2pt & 10$\times$2pt  \\
\hline

$\Omega_{\rm b}h^2$ 
& $0.0227\pm0.0017$ 
& $0.0227^{+0.0013}_{-0.0014}$ 
& $0.0227\pm0.0017$
& $0.0229\pm0.0016$ \\

$\Omega_{\rm c}h^2$ 
& $0.1212\pm0.0046$
& $0.1215^{+0.0030}_{-0.0035}$
& $0.1215\pm0.0050$
& $0.1216\pm0.0044$ \\

$n_s$ 
& $0.9592\pm0.0080$
& $0.9587\pm0.0057$
& $0.9588\pm0.0089$
& $0.9584\pm0.0082$ \\

$H_0$ 
& $67.2\pm1.5$
& $67.3^{+1.0}_{-1.1}$
& $67.3\pm1.8$
& $67.2\pm1.5$ \\

$\log(10^{10}A_s)$ 
& $3.054\pm0.028$
& $3.052\pm0.020$
& $3.051\pm0.034$
& $3.052\pm0.030$ \\

{$10^{3} a_0$} 
& $-7.588 \pm 0.026$
& $-7.588 \pm 0.022$
& $-7.588 \pm 0.035$
& $-7.587 \pm 0.033$ \\

{$10^{3} a_1$} 
& $2.008 \pm 0.036$
& $2.009 \pm 0.038$
& $2.008 \pm 0.036$
& $2.006 \pm 0.037$ \\

{$10^{3} a_2$} 
& $-4.128 \pm 0.057$
& $-4.127 \pm 0.057$
& $-4.126 \pm 0.057$
& $-4.129 \pm 0.054$ \\

{$10^{3} a_3$} 
& $2.918 \pm 0.037$
& $2.917 \pm 0.036$
& $2.917 \pm 0.037$
& $2.917 \pm 0.037$ \\

{$10^{3} a_4$} 
& $-0.678 \pm 0.011$
& $-0.678 \pm 0.011$
& $-0.678 \pm 0.011$
& $-0.678 \pm 0.011$ \\

$b_0$
& $0.8308\pm0.0033$
& $0.8304\pm0.0020$
& $0.8305\pm0.0068$
& $0.8304\pm0.0061$ \\

$b_1$
& $1.1902\pm0.0079$
& $1.1902\pm0.0068$
& $1.1907\pm0.0082$
& $1.1912\pm0.0069$ \\

$b_2$
& $-0.9281\pm0.0085$
& $-0.9279\pm0.0076$
& $-0.9283\pm0.0093$
& $-0.9288\pm0.0084$ \\

$b_3$
& $0.4232\pm0.0026$
& $0.4231\pm0.0023$
& $0.4233\pm0.0030$
& $0.4234\pm0.0027$ \\

$b_0^{\rm GW}$
& --
& $0.830\pm0.028$
& --
& $0.830\pm0.028$ \\

$b_1^{\rm GW}$
& --
& $1.191\pm0.079$
& --
& $1.193\pm0.080$ \\

$b_2^{\rm GW}$
& --
& $-0.929\pm0.063$
& --
& $-0.931\pm0.064$ \\

$b_3^{\rm GW}$
& --
& $0.424\pm0.015$
& --
& $0.424\pm0.015$ \\

$\mu_0$
& --
& --
& $0.003\pm0.059$
& $0.005\pm0.047$ \\

$\Sigma_0$
& --
& --
& $0.001\pm0.018$
& $-0.003^{+0.013}_{-0.012}$ \\

$\Omega_m$
& $0.3200\pm0.0014$
& $0.31981\pm0.00065$
& $0.3196\pm0.0050$
& $0.3209^{+0.0039}_{-0.0026}$ \\

$\sigma_8$
& $0.8154\pm0.0017$
& $0.81555\pm0.00081$
& $0.8153\pm0.0018$
& $0.8152\pm0.0011$ \\

\hline
\hline
\end{tabular}
\caption{Summary of the marginalized 68\% constraints for the four analysis configurations: $\Lambda$CDM with LSS only (3 $\times$ 2 pt) and LSS with GW probes (10 $\times$ 2 pt), $\mu\Sigma$CDM with LSS only (3 $\times$ 2 pt) and LSS with GW probes (10 $\times$ 2 pt).}
\label{tab:Results}
\end{table*}

\section{Conclusions}\label{sec:conclusions}
In this work we have investigated the potential of combining Gravitational-Wave (GWs) probes, in particular GW Number Counts (GW-NC) and GW Weak Lensing (GW-WL), with traditional Large-Scale Structure (LSS) observables from galaxies, photometric Galaxy Clustering (GCph) and galaxy Weak Lensing (WL), including all auto- and cross-correlation, to form what we referred to as 10 $\times$ 2 pt statistics. We set up a sophisticated pipeline, which comprises the creation of mock tomographic galaxies and GW data, the production of theoretical predictions for all spectra in the 10 $\times$ 2 pt analysis in the context of modified gravity and a full sampling of the posterior. In the main analysis, we did not include any relativistic corrections to galaxy and GW clustering; however, the latter are expected to provide additional constraints on MG parameters and we explore this in an appendix, via the less computationally expensive Fisher formalism.

We simulated data in the form of angular power spectra for Stage-IV LSS surveys, such as \textit{Euclid}, and future third-generation GW detectors, as Einstein Telescope (ET) and Cosmic Explorer (CE), using a modified version of Boltzmann Solver \texttt{MGCAMB} that includes the relevant GW source terms. We performed a preliminary analysis of the GW configuration, and adopted a realistic, though optimistic, setup consisting of $10^6$ events, a scale cut at $\ell_{\max}=200$, and a luminosity-distance uncertainty of $\sigma_{d_L}=1\%$. We further assumed that redshift information is available for the GW events, either through electromagnetic counterparts or through statistical galaxy-association techniques. The posterior distributions for both cosmological and nuisance parameters were sampled using \texttt{Nautilus}, interfaced with a custom likelihood implemented in \texttt{Cobaya}. All analyses include observational systematics, galaxy and gravitational bias, and Intrinsic Alignments (IA), while keeping only the dominant physical contributions to the signal being the denisty term for the clustering and the convergence one for the lensing.

We explored two cosmological scenarios: the baseline $\Lambda$CDM model, and an extended modified-gravity model, $\mu\Sigma$CDM, performing forecasts both with  LSS-only (3 $\times$ 2 pt) and  the full LSS$\times$GW combination (10 $\times$ 2 pt).
For $\Lambda$CDM, we find that the addition of GW observables leads to a clear tightening of the constraints, most notably for $\sigma_8$, one of the primary parameters controlling the amplitude of the LSS. 
The improvement is driven by the ability of GW probes to break degeneracies between cosmology and galaxy-specific systematics: GW-NC and GW-WL depend on the same underlying physics as GCph and WL, but are affected by different biases. As a result, the degeneracies between $\sigma_8$ and the nuisance parameters of GCph $b_0$ and galaxy WL $a_0$ are partially broken when GW information is included.

 For the $\mu\Sigma$CDM model, the situation is more complex. The two additional modified-gravity parameters $(\mu_0,\Sigma_0)$ partly behave as amplitude-like modifications to the clustering and lensing signals, similarly to the nuisance parameters. These parameters are then degenerate with the existing nuisance parameters $b_0$ and $a_0$, effectively reintroducing the same type of degeneracy that GWs help to break in the $\Lambda$CDM case. As a consequence, the improvement in the cosmological constraints is notably weaker. 

Overall, our results show that, in the future, GWs can significantly enhance the cosmological constraining power of LSS surveys, especially for parameters that are strongly degenerate with galaxy systematics, as in the case of $\sigma_8$ in $\Lambda$CDM. However, when moving to extended models such as $\mu\Sigma$CDM, where additional amplitude-like parameters enter the analysis, a substantially larger number of GW events would be needed for GW-NC to have a comparable impact on the clustering sector. In this context, the dominant GW 
contribution arises from GW-WL rather than by GW-NC. This should be interpreted in light of the adopted data 
configuration: given the significantly smaller number of GW events compared to galaxies, the statistical power of GW-NC is weaker. The relative importance of GW-WL is therefore fully consistent with the characteristics of the data considered here and represents the expected behaviour under these assumptions.

Looking ahead, several avenues remain open for further development. In this first study we have adopted an idealised setting, ignoring a number of instrumental and observational calibration systematics that will inevitably affect both LSS and GW measurements. On the LSS side, these include photometric-redshift uncertainties, shear calibration biases, and spatially varying systematics; on the GW side, distance-calibration errors, detector sensitivity anisotropies, and selection effects. Incorporating these effects in a fully consistent manner will be essential for realistic survey forecasts. Likewise, extending our modelling to include the full set of relativistic corrections in GW-NC, beyond the leading terms considered here, will be key to unlocking the true complementarity between GW and traditional LSS probes, especially in extended cosmologies. The effects of modified gravity upon the GW luminosity distances could also be taken into account. Future work will also explore more sophisticated data combinations, improved tomographic strategies, and the impact of larger GW event samples expected from the full network of upcoming detectors. All codes developed for this analysis, including the modified Boltzmann solver, likelihood modules, and forecasting pipeline, are publicly available at \url{https://github.com/chiaradeleo1}.
\newpage

\appendix

\section{Gravitational Waves source terms} \label{app:GW_sources}
We include here the complete list of source terms entering GW-WL and GW-NC, and that we have implemented in \texttt{MGCAMB}\footnote{\url{https://github.com/sfu-cosmo/MGCAMB.git}}. As stated in the main body, it is possible to write a general expression for the sources of the window function, as:
\begin{equation}
    S_{\rm obs}^i(k) = \int^{\eta_0}_0 d\eta\, W^i(\eta)\mathcal{S}{\rm src}(\eta,k)\, j_\ell(kd_C).
    \label{eq:sources_GW_app}
\end{equation}
In the case of GW-WL the different contributions are~\citep{Garoffolo_2021}:
\begin{itemize}
\item Convergence: $\mathcal{S}_\kappa(\eta) = [\Phi(k,\eta) + \Psi(k,\eta)] \, \displaystyle\int_0^\eta d\eta' W(\eta') \frac{\ell (\ell + 1)}{2} \frac{d_C' - d_C}{d_C' d_C} \, ,$

\item Volume: $\mathcal{S}_{\rm{vol}}(\eta) = -[\Phi(k,\eta) + \Psi(k,\eta)] \, W(\eta), $

\item Time-Delay: $\mathcal{S}_{\rm{TD}}(\eta)=[\Phi(k,\eta)+\Psi(k,\eta)] \, \displaystyle\int^\eta_0 d\eta'  \frac{W^i(\eta')}{d_C'} $,

\item Doppler: $\mathcal{S}_{\rm{Doppler}}(\eta)=-\frac{d}{d\eta}\Biggl[ \frac{W^i(\eta)}{k}\, (1-\frac{1}{\mathcal{H} d_C}) \,v(k,\eta) \Biggr]$,

\item Sachs-Wolfe: $\mathcal{S}_{\rm{SW}}(\eta) =  W(\eta )\,\frac{1}{\mathcal{H} d_C}\, \Psi(k,\eta)$,

\item Integrated Sachs-Wolfe: $\mathcal{S}_{\rm{ISW}}(\eta) = [\dot{\Phi}(k,\eta)+\dot{\Psi}(k,\eta)] \, \displaystyle\int^{\eta_0}_0 d\eta'\, W(\eta') \, (1-\frac{1}{\mathcal{H}})$,

\end{itemize}

\begin{figure}
    \centering
    \includegraphics[width=1\linewidth]{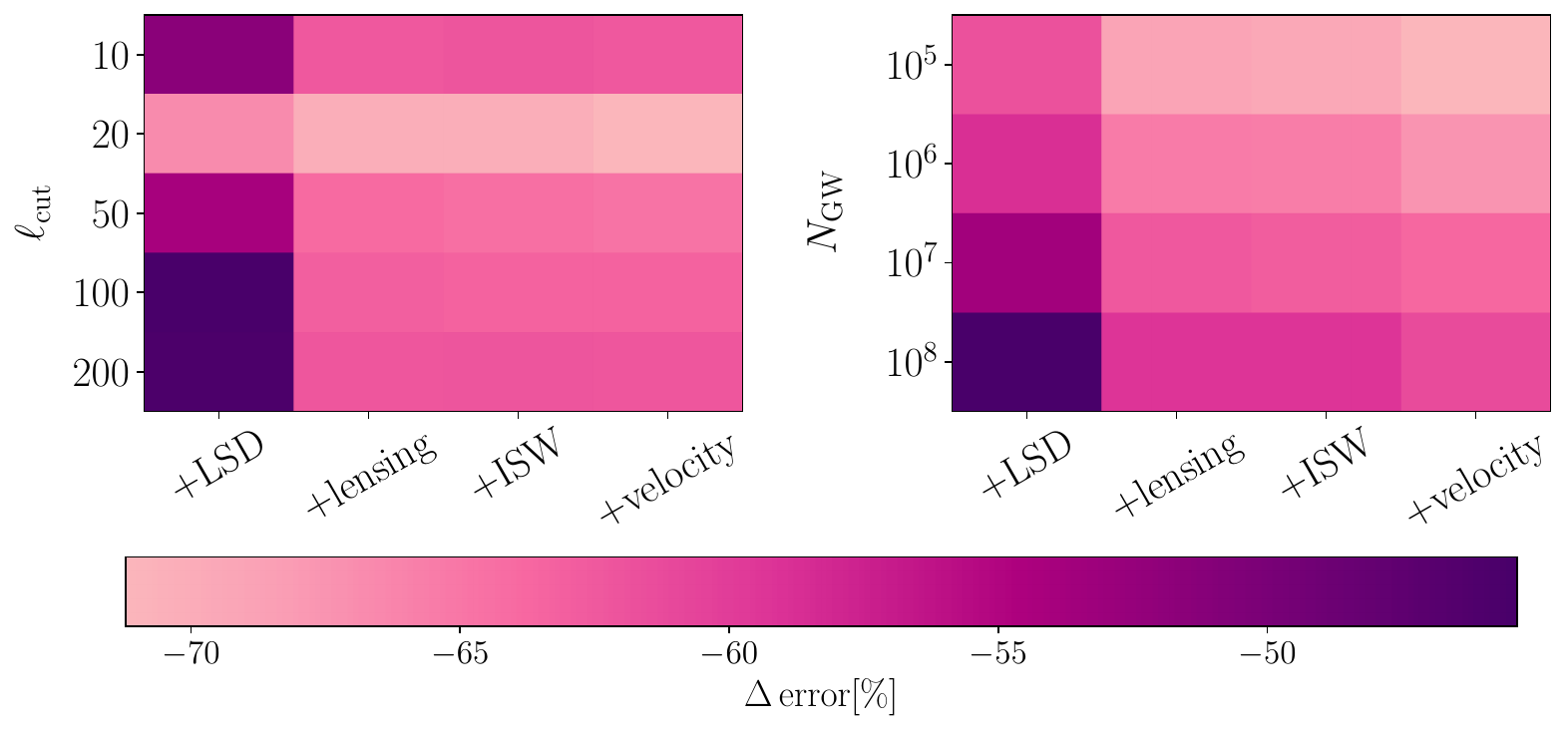}
    \caption{Relative change in the forecast error on the parameter $\mu_0$ when including relativistic effects for the GW-NC. The relative difference in forecast error is computed with respect to the baseline case that includes only density perturbations. On the x-axis, different combinations of relativistic effects are shown, where each label represents the cumulative inclusion of effects. The change in the error is shown in percentage.}
    \label{fig:error_matrix_rel_eff}
\end{figure}

In the case of GW-NC, following the definitions introduced in \citep{Balaudo_2024}, we define the functions: 
\begin{equation}
\gamma=\Biggl[1+\frac{1}{\Bar{d_C}\mathcal{H}}-\frac{\alpha_M}{2}\Biggr]^{-1}, 
\label{eq:gamma}
\end{equation} 
and 
\begin{equation} 
\beta=\gamma \Biggl[\frac{2}{\Bar{d_C}\mathcal{H}}+\frac{\dot{\mathcal{H}}}{\mathcal{H}^2}-\frac{\gamma}{\Bar{d_C}\mathcal{H}}\Biggl(1+\frac{\dot{\mathcal{H}}}{\mathcal{H}^2}\Biggr)-\frac{\alpha_M}{2}\Biggl(1+\frac{2}{\Bar{d_C}\mathcal{H}}\Biggr)+\frac{\alpha_M^2}{4}+\frac{\dot{\alpha_M}}{2\mathcal{H}}\Biggr], 
\label{eq:beta}
\end{equation} 
which encapsulate the modifications to the propagation of GWs due to an evolving Planck mass and the background expansion. With these definitions, the contributions to the GW-NC observable can be expressed as follows:
\begin{itemize}
\item Density: $\mathcal{S}_{\rm{dens}}(\eta)=W^i(\eta) \delta_{GW} (k,\eta)$,

\item Time-Delay: $\mathcal{S}_{\rm{TD}}(\eta)=[\Phi(k,\eta)+\Psi(k,\eta)]\,\displaystyle\int^\eta_0 d\eta'\, W^i(\eta')\, (\frac{1-\beta}{d_C'} + \frac{\gamma}{\mathcal{H}d_C'^2})$,

\item Doppler: $\mathcal{S}_{\rm{Doppler}}(\eta)=\dfrac{d}{d\eta}\Biggl[ \dfrac{W^i(\eta)}{k}\, (-1-2\gamma-2\beta) \,v(k,\eta) \Biggr]$,

\item Integrated Sachs-Wolfe (ISW):  $\mathcal{S}_{\rm{ISW}}(\eta) = [\dot{\Phi}(k,\eta)+\dot{\Psi}(k,\eta)]\displaystyle\int^{\eta_0}_0 d\eta'\, W(\eta')\,2(\beta+1)$,

\item Potential: $\mathcal{S}_{\Phi}(\eta)=W^i(\eta) \Biggl[(\beta-1-\frac{\gamma}{\Bar{d_C}\mathcal{H}})\Phi+\dfrac{\gamma}{\mathcal{H}}\dot{\Phi}+(2\beta+3-\dfrac{\gamma}{\Bar{d_C}\mathcal{H}}) \Biggr]$,

\item Gradient of Potential: $\mathcal{S}_{\nabla \Phi}(\eta)=\dfrac{d}{d\eta}\Biggl[ \dfrac{W^i(\eta)}{k} \,\dfrac{\gamma}{\mathcal{H}}\, \Phi(k,\eta)\Biggr]$,

\item Luminosity Space Distortions (LSD): $\mathcal{S}_{\rm{LSD}}(\eta)=\displaystyle\int^{\eta_0}_0 d\eta \,\frac{d^2}{d\eta^2} \,\Biggl[2\frac{W^i(\eta)}{k}\frac{\gamma}{\mathcal{H}}\,v(k,\eta)\Biggr]$,

\item Lensing: $\mathcal{S}_{\rm{lensing}}(\eta)=-\ell(\ell +1)\, \displaystyle\int^{\eta_0}_0 d\eta \,\displaystyle\int^{\eta}_0 d\eta' W(\eta')\,\Bigl[\frac{\beta -1}{2}  \frac{d_C'-d_C}{d_C'd_C}+\frac{\gamma}{2\mathcal{H}d_C'^2}\Bigr][\Phi(k,\eta)+\Psi(k,\eta)]$.
\end{itemize}
In \autoref{fig:error_matrix_rel_eff}, we illustrate the impact of including relativistic effects relative to the density-only case for the GW-NC, defined as 
\begin{equation}
    \Delta\rm error = \frac{\sigma^{density+rel}-\sigma^{density}}{\sigma^{density}}.
\end{equation} Each combination shows the cumulative effect of adding relativistic contributions, specifically LSD, Lensing, ISW, and Doppler, to the density-only scenario. The analysis is performed across different values of $\ell_{\rm cut}$ and varying numbers of GW events.
In the left panel, with a fixed number of GW events ($N_{\rm GW} = 10^6$), we examine the influence of scale cuts. At low multipoles, the inclusion of relativistic effects yields minimal improvement in the error, primarily due to the dominance of cosmic variance. As we move to intermediate scales ($\ell \sim 20$–$50$), we observe the most significant improvement, since these are the scales where relativistic effects become most influential. At higher multipoles, the impact of relativistic terms decreases, even if the lensing continues to provide an improvement, whereas the LSD contribution alone offers limited enhancement over the density term, which dominates at these scales. In the right panel, we investigate how the forecast error changes with the number of GW events, fixing the scale cut at $\ell_{\rm cut} = 20$. The results indicate that the relative gain from including relativistic effects is more pronounced when the number of events is low. As the number of GW events increases, the statistical power of the data improves, reducing the benefit provided by relativistic corrections.

A convenient and widely used way to characterize deviations from GR in the propagation of GWs is through the ratio between the GW and EM luminosity distances, which in scalar-tensor theories can be written as:
\begin{equation}
\Xi(z) \equiv \frac{d_L^{\rm GW}(z)}{d_L^{\rm EM}(z)} 
= \exp\!\left[ -\frac{1}{2} \int_0^z \frac{\alpha_M(\tilde z)}{1+\tilde z}\, d\tilde z \right],
\label{eq:Xi}
\end{equation}
where $\alpha_M$ parametrizes the running of the effective Planck mass and Eq.~\eqref{eq:Xi} follows from solving the GW propagation equation in scalar-tensor theories. The parameter $\alpha_M$ can be defined as:
\begin{equation}
\alpha_M \equiv \frac{d\ln M_*^2}{d\ln a}
= \frac{d\ln G_{\rm eff}^{-1}}{d\ln a},
\end{equation}
so that the modification of the GW amplitude depends on the evolution of the effective gravitational coupling along the line of sight. Using this definition, Eq.~\eqref{eq:Xi} can be rewritten in terms of the ratio of the effective gravitational coupling evaluated at emission and at detection, in the absence of additional propagation effects.
\newline
In screened scalar-tensor theories, GWs sources and the observer are located in high-density environments where screening mechanisms are efficient. Under this assumption, the effective gravitational coupling at emission and detection is well approximated by the locally measured Newton constant, implying that GW generation and detection are governed by GR. Deviations from GR therefore enter primarily through the propagation encoded in Eq.~\eqref{eq:Xi}, rather than through local effects.

To model the running of the Planck mass, we follow a phenomenological approach and express $\alpha_M$ in terms of a non-minimal coupling function $\Omega$ as:
\begin{equation}
\alpha_M(\eta) = \frac{d\Omega/d\eta}{\mathcal{H}(\eta)(1 + \Omega)}, 
\label{eq:alpha_M} 
\end{equation}

where we take $\Omega(a) = \Omega_0\,\Omega_{\rm DE}(a)$ with $\Omega_0 \ll 1$, corresponding to weakly coupled models. In this case, the smallness of $\alpha_M$ is controlled by the amplitude parameter $\Omega_0$.
With this parametrization, in GW propagation, deviations from GR are expected to be small. While we focus on regimes where $\alpha_M$ remains perturbative, its effect can accumulate over cosmological distances and lead to observable signatures in GW-based observables. Relativistic projection effects provide additional corrections to GW observables, but these are subleading contributions, and in modified-gravity scenarios they correspond to corrections to already small relativistic terms.

\section{Validation of the Intrinsic Alignment polynomial model} \label{app:IA}
\begin{figure}[t]
    \centering
    \includegraphics[width=1.1\linewidth]{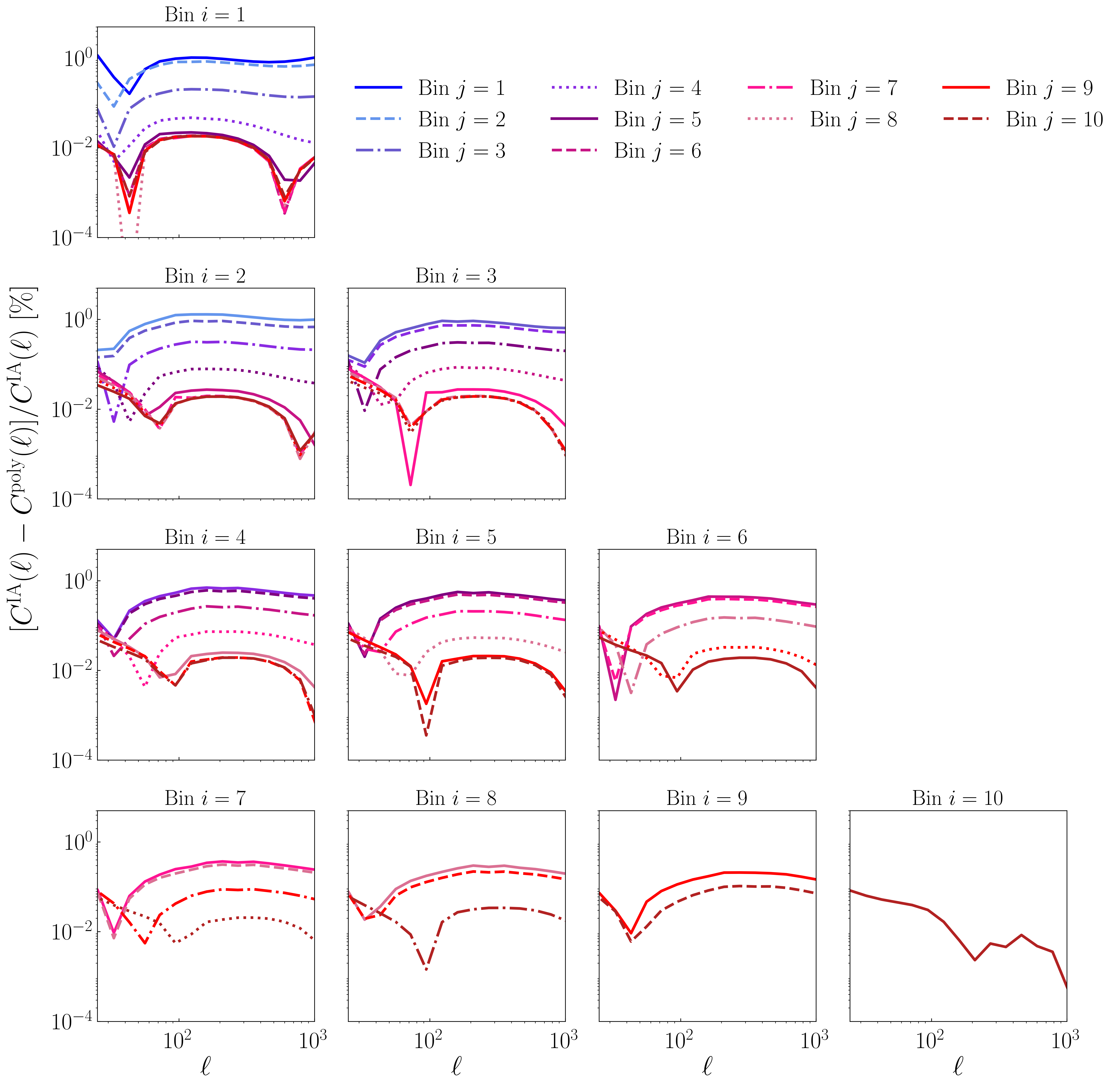}
    \caption{Relative difference in the Angular Power Spectrum coefficients evaluated with the commonly adopted expression for the IA and the polynomial one. We show the relative difference evaluated for $C^{{\rm WL}_i \times {\rm WL}_j}(\ell)$, where in each subplot the $i$-th bin is fixed, while the $j$-th varies.}
    \label{fig:IA}
\end{figure}

In this work we model the intrinsic-alignment (IA) contribution using a polynomial parametrisation in redshift,
\begin{equation}
    f_{\rm IA}(z)=a_0 + a_1 z + a_2 z^2 + a_3 z^3 + a_4 z^4,
\end{equation}
which enters the WL observables through the IA window function defined in Eq.~\eqref{window_IA}. This choice is entirely motivated by technical considerations: the current implementation of the \texttt{CAMB} source functions does not include 
IA contributions in the WL kernels.
To incorporate IA effects within our computation of angular power 
spectra, we exploit the fact that the IA window function has the same redshift dependence as the photometric galaxy clustering window function, defined in Eq.~\eqref{window_gal}, with the galaxy bias replaced by an IA amplitude. By expressing this amplitude as a polynomial function of redshift, we are able to 
implement the IA contribution as an additional window function that is fully compatible with the existing \texttt{CAMB} infrastructure and can be evaluated consistently in redshift bins.

For reference, and to guide the choice of the polynomial form, we compare this parametrisation to the commonly adopted analytical IA model:
\begin{equation}
    \mathcal{F}_{\rm IA}(z)= - \frac{A_{\rm IA} C_1 \Omega_m}{D_1(z)} (1+z)^{\eta_{\rm IA}},
    \label{FIA}
\end{equation}
where $C_1=0.00134$, $D_1(z)$ is the linear growth factor, $\eta_{\rm IA}$ controls the 
redshift dependence, and $A_{\rm IA}$ is the overall IA amplitude. The polynomial coefficients $a_i$ are obtained by fitting the redshift dependence of this reference model over the redshift range of interest. This procedure ensures that the IA signal 
included in the WL observables reproduces the expected behaviour, while remaining compatible with the \texttt{CAMB}-based implementation.

As shown in \autoref{fig:IA}, the relative difference between the polynomial and the analytic IA model is negligible, ensuring that the calculations are simplified while maintaining a good approximation of the physical effects.

\section{Pipeline validation}\label{app:validation}
\begin{figure}[t]
    \centering
    \includegraphics[width=1.0\linewidth]{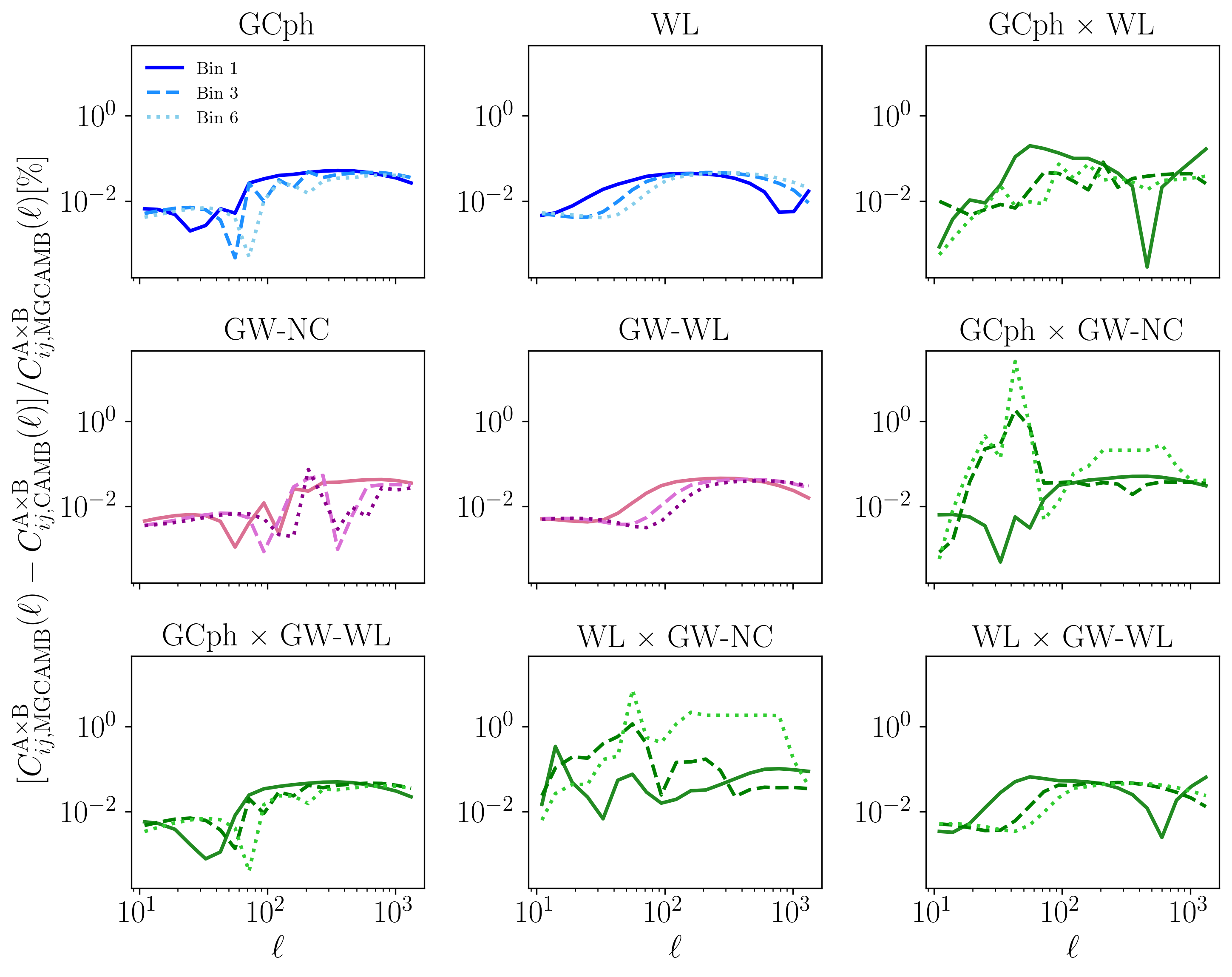}
    \caption{Relative difference in the angular power spectra coefficients for galaxy observables (blue), GW observables (purple), and the cross-correlation between the two probes (green), as computed with our version of \texttt{MGCAMB} and the private code used in \citep{Balaudo_2024}. The results are consistent, once the difference in the sign convention used in the GW-WL convergence term in the two codes is taken into account.}
    \label{fig:Code_validation}
\end{figure}

At the core of our analysis there is the implementation of the window function source terms for Gravitational Waves Weak Lensing (GW-WL) and Gravitational Waves Number Counts (GW-NC) in a customized version of \texttt{MGCAMB}\footnote{\url{https://github.com/sfu-cosmo/MGCAMB.git}}.
\noindent
The theoretical expressions for these GW window functions are taken from 
\cite{Garoffolo_2021, Balaudo_2024} and are summarized for convenience in 
\autoref{app:GW_sources}. Their implementation is carried out directly within the 
\texttt{fortran} source code of \texttt{MGCAMB}, starting from the public version 
available at commit \texttt{cb1a03e}. This extension allows us to compute all GW 
auto- and cross-angular power spectra consistently with the LSS treatments already implemented in \texttt{MGCAMB}.
This code has then been validated against the code used in \cite{Balaudo_2024}, that is not public yet.
In \autoref{fig:Code_validation}, we show the relative difference between the angular power spectra coefficients $C_{ij}^{\rm A\times B}(\ell)$ computed using \texttt{MGCAMB}  and the pipeline of \citep{Balaudo_2024}, for three redshift bins $i = \{1, 3, 6\}$. 
The results for GW-NC include only the density source term, while the GW-WL signal includes only the lensing convergence term.
The comparison demonstrates good agreement between the two codes across all probes and bin combinations. 
\clearpage

\acknowledgments

This article is based upon work from COST Action CA21136 Addressing observational tensions in cosmology with systematics and fundamental physics (CosmoVerse) supported by COST (European Cooperation in Science and Technology). CDL thanks financial support from the research grant number 2022E2J4RK "PANTHEON:
Perspectives in Astroparticle and Neutrino THEory with Old and New messengers"
under the program PRIN 2022 funded by the Italian Ministero dell’Universit\`a
e della Ricerca (MUR).
CDL and MM acknowledge financial support from Sapienza Università di Roma, provided through Progetti Medi 2021 (Grant No. RM12117A51D5269B). This work made use of Melodie, a computing infrastructure funded by the same project, and PLEIADI, a computing infrastructure installed and managed by INAF. 
GCH acknowledges support through the European Space Agency research fellowship programme and the ESA Science Faculty visitor research programme. GCH further acknowledges that this publication is part of the project UNICORN with file number VI.Veni.242.110 of the research programme Talent Programme Veni Science domain 2024 which is (partly) financed by the Dutch Research Council (NWO) under the grant \url{https://doi.org/10.61686/ZCPQI32997}. GCH acknowledge the EuroHPC Joint Undertaking for awarding this project access to the EuroHPC supercomputer LEONARDO, hosted by CINECA (Italy) and the LEONARDO consortium through an EuroHPC Extreme Access call.
MM acknowledges funding by the Agenzia Spaziale Italiana (\textsc{asi}) under agreement n. 2024-10-HH.0 and support from INFN/Euclid Sezione di Roma. AB and TB are supported by ERC Starting Grant SHADE (grant no.\,StG 949572). TB is further supported by a Royal Society University Research Fellowship (grant no.URF\textbackslash R\textbackslash 231006). AS acknowledges support from the European Research Council under the H2020 ERC Consolidator Grant “Gravitational Physics from the Universe Large scales Evolution” (Grant No. 101126217 — GraviPULSE)
\bibliographystyle{JHEP}
\bibliography{biblio.bib}

\end{document}